\documentclass[journal]{IEEEtran}

\usepackage{array}
\usepackage{cite}
\usepackage[pdftex]{graphicx}
\usepackage{changes}
\usepackage{amsmath}

\usepackage{algorithmic}
\usepackage{url}
\usepackage{caption}
\usepackage{subcaption}
\usepackage[capitalise]{cleveref}
\usepackage{multirow}
\usepackage{verbatim}

\usepackage[acronym,nomain,nonumberlist,nopostdot]{glossaries}
\usepackage{upgreek}
\usepackage{booktabs}
\usepackage{arydshln}

\usepackage{url}
\usepackage{tikz}
\usetikzlibrary{calc}

\newcommand{\positiontextbox}[4][]{%
	\begin{tikzpicture}[remember picture,overlay]
		\node[inner sep=3pt, fill=yellow,align=left,draw,line width=1pt,#1] at ($(current page.north west) + (#2,-#3)$) {\parbox{.80\paperwidth}{#4}};
	\end{tikzpicture}%
}

\newacronym{ADC}{ADC}{analog-to-digital converter}
\newacronym{ANN}{ANN}{artificial neural network}
\newacronym{AOA}{AOA}{angle of arrival}
\newacronym{AOD}{AOD}{angle of departure}
\newacronym{AP}{AP}{access point}
\newacronym{BLE}{BLE}{Bluetooth/Bluetooth low energy}
\newacronym{CAGR}{CAGR}{compound annual growth rate}
\newacronym{CDF}{CDF}{cumulative distribution function}
\newacronym{CITIC}{CITIC}{Centre for Information and Communications Technology Research}
\newacronym{CIR}{CIR}{channel impulse response}
\newacronym{DAC}{DAC}{digital-to-analog converter}
\newacronym{ECDF}{ECDF}{empirical cumulative distribution function}
\newacronym{FTM}{FTM}{fine time measurement}
\newacronym{GNSS}{GNSS}{global navigation satellite system}
\newacronym{GP}{GP}{Gaussian process}
\newacronym{GPIO}{GPIO}{general-purpose input/output}
\newacronym{I2C}{I$^2$C}{Inter-Integrated Circuit}
\newacronym{I2S}{I$^2$S}{Inter-IC Sound}
\newacronym{INS}{INS}{inertial navigation system}
\newacronym{IOT}{IoT}{internet of things}
\newacronym{IR}{IR}{infrared}
\newacronym{LBS}{LBS}{location based services}
\newacronym{LCD}{LCD}{liquid-crystal display}
\newacronym{LDO}{LDO}{low-dropout regulator}
\newacronym{LED}{LED}{light-emitting diode}
\newacronym{LOS}{LoS}{line of sight}
\newacronym{MCU}{MCU}{microcontroller unit}
\newacronym{ML}{ML}{machine learning}
\newacronym{NLOS}{NLOS}{non line of sight}
\newacronym{NN}{NN}{neural network}
\newacronym{PWM}{PWM}{pulse-width modulation}
\newacronym{ReLU}{ReLU}{rectified linear unit}
\newacronym{RF}{RF}{radio frequency}
\newacronym{RFID}{RFID}{radio frequency identification}
\newacronym{ROM}{ROM}{read-only memory}
\newacronym{ROS}{ROS}{Robot Operating System}
\newacronym{RSS}{RSS}{received signal strength}
\newacronym{RSSI}{RSSI}{received signal strength indicator}
\newacronym{RTT}{RTT}{round trip time}
\newacronym{SRAM}{SRAM}{static random-access memory}
\newacronym{SPI}{SPI}{serial peripheral interface}
\newacronym{SOC}{SOC}{system on a chip}
\newacronym{SVM}{SVM}{support vector machine}
\newacronym{TDOA}{TDOA}{time difference of arrival}
\newacronym{TOA}{TOA}{time of arrival}
\newacronym{TOF}{TOF}{time of flight}
\newacronym{UART}{UART}{universal asynchronous receiver-transmitter}
\newacronym{UWB}{UWB}{ultra wideband}

\begin{document}
%
\title{Fine Time Measurement for the Internet of Things: A Practical Approach Using ESP32}

\author{Valentín Barral,
Omar Campos,
Tomás Domínguez-Bolaño,
Carlos J. Escudero,~\IEEEmembership{Member,~IEEE}, and
José A. García-Naya,~\IEEEmembership{Senior Member,~IEEE}

\thanks{Manuscript received 20 September 2021 and 9 February 2022: accepted 26 February 2022. Date of publication 11 March 2022. This work has supported in part by the Galician Innovation Agency (GAIN) and the Regional Ministry of Economy, Employment and Industry, Xunta de Galicia under Grant COV20/00604; in part by Xunta de Galicia under Grant ED431C 2020/15 and Grant ED431G 2019/01 to support the Centro de Investigación de Galicia ``CITIC''; in part by the Agencia Estatal de Investigación of Spain under Grants RED2018-102668-T and Grant PID2019-104958RB-C42; and in part by the European Regional Development Fund (ERDF) Funds of the EU (FEDER Galicia 2014-2020 and AEI/FEDER Programs, UE). \textit{(Corresponding author: Valentín Barral Vales.)} } %
\thanks{The authors are with the CITIC Research Center, Department of Computer Engineering, University of A Coruña, 15071 A Coruña, Spain. (e-mail: \mbox{valentin.barral@udc.es}, \mbox{omar.campos@udc.es}, \mbox{tomas.bolano@udc.es}, \mbox{escudero@udc.es}, and \mbox{jagarcia@udc.es})}
\thanks{Digital Object Identifier 10.1109/JIOT.2022.3158701.}
}

\markboth{IEEE Internet Of Things Journal,~VOL.~9, NO.~19, 1~OCTOBER~2022}{}

\maketitle

\begin{abstract}
In the world of internet of things (IoT), obtaining the physical location of devices has always been a task of great interest for developing increasingly complex location-based services (LBS). That is why in recent years wireless communication standards have been incorporating new additions focused on providing localization mechanisms to technologies widely used in the IoT world, such as Wi-Fi or Bluetooth. In particular, the IEEE 802.11-2016 Wi-Fi standard introduced ranging estimation between two devices through the so-called fine time measurement (FTM) protocol, defined by the IEEE 802.11mc. FTM is not yet widespread in the IoT field, but commercial modules capable of offering this functionality at a reasonable price are starting to appear. In early 2021, the most widespread system on a chip (SOC) family among IoT devices, the ESP32-XX series, added support for this Wi-Fi standard, enabling, for the first time, the use of a standard designed for location-based systems. This paper analyzes the performance of this FTM implementation by carrying out and studying several measurement campaigns in different indoor and outdoor scenarios. Additionally, this work proposes an alternative real-time implementation for distance estimation inside the ESP32 using an approach based on machine learning. Such an implementation is successfully validated in a scenario totally different than those considered for the training and test sets. Finally, both the measurement sets and the developed software are available to the scientific community.
\end{abstract}


\IEEEpeerreviewmaketitle
\positiontextbox{11cm}{27cm}{\footnotesize This work is licensed under a Creative Commons Attribution 4.0 License. For more information, see \url{https://creativecommons.org/licenses/by/4.0/}. \\ The final published version of this work is available open access, with the same license, at \url{https://doi.org/10.1109/JIOT.2022.3158701}}


\section{Introduction}

Accurate location of sensors in the \gls{IOT} world is important in multiple areas of interest such as smart cities and buildings, healthcare environments, industry, or agriculture \cite{khelifiSurveyLocalizationSystems2019}. All of them demand location technologies capable of positioning the sensors to deliver more and better \gls{LBS}. Recently, the global pandemic of COVID-19 demonstrated the important need not only to know all kinds of data, such as CO$_2$ rates or occupancy levels, but also to know the physical location of the sensors that produce them \cite{tedeschiIoTraceFlexibleEfficient2021}. The location of sensors within the aforementioned scenarios often becomes challenging due to the following reasons \cite{liLocationEnabledIoTLEIoT2021}:

\begin{itemize}
    \item In many occasions, \glspl{GNSS} are not an option because the sensors are located in scenarios (inside buildings, urban canyons, etc.) where the performance of such systems is very poor or they simply do not work. And even in open areas where these systems could work, the cost of adding such capabilities to the sensors would be too high to be viable for large-scale deployments.
    \item The complexity of the scenarios in which \gls{IOT} sensors are placed often makes their localization very difficult. This is especially true when \gls{RF} technologies are employed. Phenomena such as interference or \gls{NLOS} propagation yield significant errors in position estimates.
    \item Typical deployments in the \gls{IOT} world require the installation of many sensors, which means that the individual cost of each unit must also be very low in order not to exceed the project budget. Therefore, in certain cases, some localization technologies are excluded in advance for budget reasons, even if they perform satisfactorily.
\end{itemize}

There are, therefore, numerous technological alternatives to achieve position estimation within the \gls{IOT} world, each with its own particular characteristics in terms of accuracy level, deployment and maintenance costs, reliability, or operational requirements. 

Among the various localization systems available \cite{brena_evolution_2017,sakpere_state---art_2017,laoudias_survey_2018}, those based on \gls{RF} are some of the most popular. Their basic operation consists in emitting and receiving some type of waveform with certain characteristics and extracting from this transmission some physical parameters that allow for estimating the position of the emitter or receiver. 

Based on these RF-based technologies there are numerous implementations, including \gls{RFID} \cite{seco_smartphone-based_2018}, Wi-Fi (IEEE 802.11) \cite{yang_wifi-based_2015}, \gls{BLE} \cite{teran_iot-based_2017}, Zigbee \cite{huang_zigbee-based_2011}, and \gls{UWB} \cite{sahinoglu_ultra-wideband_2008}.

However, when applications require very accurate and precise localization, the range of technologies that can be considered is restricted \cite{brena_evolution_2017}. Traditional technologies exclusively provide \gls{RSS} estimates, which can be heavily affected by the signal propagation conditions, and thus they may be highly variable and unpredictable, especially in indoor environments. Therefore, in recent years, newer technologies providing parameters such as \gls{TOA}, \gls{TDOA}, \gls{AOA}, or \gls{AOD} are considered since they exhibit a much better accuracy and precision than those based on \gls{RSS}. Many of the latest versions of \gls{RF}-based wireless communication standards offer these new alternatives since they are evolving with the aim of providing reliable references for application in location systems. In particular, Wi-Fi (with amendments IEEE 802.11az/bd/mc) \cite{au_latest_2016, hashem_accurate_2021, gentner_wifi-rtt_2020} and 5G \cite{dammann_prospects_2015, buehrer_collaborative_2018} provide \gls{TOA} and \gls{TDOA}, whereas Bluetooth 5.1 with direction finding \cite{suryavanshi_direction_2019} provides \gls{AOA} and \gls{AOD}.

Wi-Fi networks are one of the widely used mechanisms for communication between \gls{IOT} nodes (e.g., in home automation). In previous years, the Task Group mc (TGmc) of the IEEE 802.11 Working Group (IEEE 802.11mc), incorporated several amendments into the IEEE 802.11 standards, resulting in the publication of the IEEE 801.11-2016 standard in 2016. This standard incorporates a new protocol for estimating the propagation time between devices, the so-called \gls{FTM} protocol. The advantage of this standard is that it has been introduced in recent years into consumer devices and, for example, is supported by the Android operating system since version 9.0 (Android Pie with API level 28) \cite{noauthor_wi-fi_nodate}. In fact, some tests with commercial smartphones using this standard are already available in the literature \cite{bullmann2020comparison, horn2020doubling, ibrahim2018verification}.

The same is not true for the implementation of this technology in devices employed in the \gls{IOT} world. In this case, there are very few choices offering Wi-Fi \gls{FTM} off-the-shelf. One of the main reasons is that most Wi-Fi modules on the market are not yet compatible with the 802.11-2016 standard, and only a few recent models, such as the 88W8987 family from NXP \cite{88W8987DB1x1}, offer this functionality. It is important to mention that these modules do not offer an autonomous solution since they need an external micro-controller to act as a host and communicate with them. However, there is another alternative, which are the \gls{SOC} devices that already incorporate a micro-controller (typically a low-power model), and a series of additional communication modules such as Wi-Fi or Bluetooth. Several chips of this type are available off-the-shelf, such as the 88MW32X from NXP \cite{88MW32XWiFiMicrocontroller}, or the CYW43907, CYW43903, and CYW54907 from Cypress \cite{WirelessMCUs}. In all these models, however, the Wi-Fi module built into the \gls{SOC} does not support the 802.11-2016 standard. Currently, to the best of our knowledge, the only option available on the market with \gls{FTM} support is the well known ESP32-XX family of chips from the Chinese manufacturer Espressif Systems. Specifically, its ESP32-S2 model released in late 2019 was the first to support \gls{FTM}, with the release in February 2021 of its firmware version 4.3-Beta1 \cite{ReleaseESPIDFPrerelease}.

It is more than likely that this will lead to an explosion of \gls{LBS} in the \gls{IOT} world due to the high popularity and market penetration of these \gls{SOC} models. It is important to know that the \gls{SOC} market is projected to grow from USD 471 millions in 2020 to USD 1656 millions in 2025 with a \gls{CAGR} of 28.6\,\% \cite{noauthor_global_nodate}.

In addition, we should keep in mind that the consumer device industry and communities are focused on the development of devices based on the ESP32-XX family for multiple reasons: low cost (around 5\,USD each), low energy consumption, compatibility with popular development environments (Arduino, MicroPython, etc.), and the availability of high-quality documentation. Thus, a wide variety of applications have emerged in various fields (industry, home automation, monitoring, security, wearables, surveillance, location, traceability, etc.). Another example of its success is that there is a wide variety of firmware versions that allow these \glspl{SOC} to be integrated in various automation projects, for example, ESPHome \cite{noauthor_esphome_nodate}, Tasmota \cite{noauthor_tasmota_nodate}, or ESP easy \cite{noauthor_esp_nodate}.

This paper focuses on assessing whether the existing \gls{FTM} implementation in a module, such as the ESP32-S2, is mature enough in terms of accuracy and robustness to build large-scale \gls{LBS} in the \gls{IOT} world. For this purpose, several measurement campaigns were carried out in realistic and heterogeneous environments, including indoors and outdoors, recording the distance values estimated by the chip and comparing them with the actual values. The dataset of such measurements is publicly available in \cite{2pv8-ze59-21}. 

During this performance study, a large difference in the distance estimation from the raw time values was observed depending on the measuring scenario: indoors or outdoors. Therefore, taking advantage of the ease of programming the ESP32, this paper presents another contribution: the proposal of an alternative method to estimate the distance from the raw time values returned by the Wi-Fi module. Such a proposal is based on the use of \gls{ML} algorithms trained with part of the acquired data during the measurements. The performance of this alternative is tested not only with offline measurements, but also with a real-time implementation running on the ESP32-S2 chip. Moreover, such a real-time implementation is validated in a scenario totally different than those considered for the training and test sets. Finally, the code of such an implementation was also publicly released under an open source license (see \cite{esp32ftmtag, esp32ftmanchor, rosftmnode, rosftmmsgs}).

The remainder of this paper is organized as follows. \cref{sec:related_work} includes an analysis of the state of the art related to Wi-Fi \gls{FTM} measurements. \cref{sec:measurements} describes the ESP32-S2 and the measurement scenarios. The obtained results are analyzed in \cref{sec:measurement_analysis}, whereas \cref{sec:ml} presents an approach based on \gls{ML} to obtain distance estimates considering both \gls{RTT} and \gls{RSSI}. A description of a new testing scenario to check the robustness of the estimation using \gls{ML} is included in \cref{sec:test} together with the results obtained after the experiments. \cref{sec:test_deployment} details the implementation of the estimator built within the ESP32-S2, as well as an analysis of the current consumption when performing \gls{FTM} operations. Finally, \cref{sec:conclusions} is devoted to the conclusions and future work.

\section{Related Work}\label{sec:related_work}
Since the definition of the \gls{FTM} protocol in the IEEE 802.11-2016, several works have focused on analyzing its performance in distance and location estimation tasks.

In \cite{ibrahim2018verification}, a study is presented on the performance of several \gls{FTM}-capable Wi-Fi chips in different indoor and outdoor scenarios, with and without \gls{LOS} propagation. The study considered 20\,MHz and 40\,MHz bandwidths at the 2.4\,GHz frequency band, and a bandwidth of 80\,MHz at the 5\,GHz frequency band. The results showed an accuracy about one meter in open space (after correction for a constant offset), and about five meters in indoors. This work also showed little difference between using 20\,MHz or\,40 MHz bandwidths at the 2.4\,GHz band. However, there was a clear improvement when using the 80\,MHz bandwidth in the 5\,GHz band.

In \cite{dvorecki2019machine}, the authors considered \gls{ML} to train an \gls{ANN} capable of predicting the \gls{TOA} of the first path in a Wi-Fi signal with the \gls{FTM} protocol. In this case, \gls{CIR} samples are employed, achieving an estimator that yielded an error below 4\,m for 90\% of the samples analyzed.

In \cite{choi2019unsupervised}, different neural networks were applied to perform position estimation using \gls{RSS} and \gls{FTM} on different smartphones. Considering the 5\,GHz band and a bandwidth of 40\,MHz, measurements in an indoor environment showed that only 20\,\% of the samples were below 4\,m of error. These values increased to almost 90\,\% of the samples below 4\,m of error when an \gls{ANN}-based correction factor was applied to the raw samples. A comparison between position accuracy using \gls{RSSI} or \gls{FTM} was performed in \cite{bullmann2020comparison}. The measurements were carried out in the 2.4\,GHz band and with a bandwidth of 20\,MHz. The findings indicated that the \gls{RSSI} estimates outperform those obtained from the \gls{FTM} sensors, unless the sensors are calibrated individually, in which case the \gls{FTM} localization is the best. The error values in this case yielded a mean error of 3.52\,m for calibrated \gls{FTM} and a mean error of 4.47\,m using \gls{RSSI}. 
In \cite{horn2020doubling}, the author used a smartphone to carry out Wi-Fi \gls{FTM} measurements to obtain a positioning estimator. To improve its accuracy, the author integrated the measurements from two different frequency bands, 2.4\,GHz and 5\,GHz, showing that there is no correlation between them. The work also included a study of the possible reasons why errors appear with this technology.

In general, all the aforementioned works showed that the Wi-Fi \gls{FTM} technology can provide good results in terms of accuracy in outdoor environments without obstacles, but many problems arise in indoor environments. In these complex scenarios, the studies showed that, without a minimum calibration or optimization of each device, the results may lead to errors of tens of meters. In all the studies, Wi-Fi cards or smartphones were used as initiating devices, hence none of them addressed the problem of energy consumption or cost when describing the technology. In this paper, we will analyze whether the ESP32-S2 offers results with similar accuracy to those obtained in previous works, but also taking into account its characteristics as a device to be used in \gls{IOT} environments.

\section{Experimental Setup}\label{sec:measurements}
\subsection{Fine Time Measurements and the ESP32}\label{sec:ftm_esp32}
The ESP32-S2 \cite{espresssif_systems_esp32-s2_2021} is an evolution of the famous ESP8266, massively used in automation projects, which introduces new Bluetooth and Wi-Fi capabilities. Both ESP32-S2 and ESP8266 \glspl{SOC} are part of a family of low-cost \gls{SOC} microcontrollers from Espressif Systems, a Chinese company from Shanghai.

In particular, the ESP32-S2 and the ESP32-C3 series have recently added support for the IEEE 802.11-2016 standard, which introduces the \gls{FTM} protocol to provide ranging estimates using Wi-Fi links. Basically, this protocol measures the \gls{RTT} between two devices and, based on this time, estimates the distance between them. These devices can be of any type (\glspl{AP}, mobile terminals, \glspl{SOC}, etc.), provided that they have a Wi-Fi chipset compatible with the 802.11-2016 standard and a firmware that supports \gls{FTM}. Due to its enormous market penetration, this work focuses on the ESP32-S2, although the results should be similar for the ESP32-C3.

The ESP32-S2 \cite{espresssif_systems_esp32-s2_2021} has integrated support for the IEEE 802.11 b/g/n and a variety of peripherals. It has 43 \gls{GPIO} pins, a 240\,MHz Xtensa 32-bit LX7 single-core processor with 320\,KB of \gls{SRAM}, and 128\,KB of \gls{ROM}. Compared to its predecessor, the ESP32-S2 \gls{SOC} comes with enhanced encryption capabilities and improved radio performance. There are also versions of the \gls{SOC} in the form of modules, such as the ESP32-S2-WROOM and the ESP32-S2-WROVER; and development kits such as the ESP32-S2-DevKitM-1 or the ESP32-S2-Saola-1. Using the 43 \gls{GPIO} pins, the device can be configured to provide different peripheral interfaces, including four \glspl{SPI}, two \gls{I2S} and \gls{I2C} interfaces, three \gls{UART} interfaces, 16 \gls{PWM} channels for \glspl{LED}, a \gls{LCD} interface, a camera interface, 18 \gls{ADC} channels, two 8-bit \gls{DAC} channels, and 14 capacitive touch interfaces.

The most important characteristic of the ESP32-S2 for this work is that it supports the IEEE 802.11-2016 standard with the \gls{FTM} protocol, which enables localization based on distance estimates between two Wi-Fi devices. However, it should be noted that only the 2.4\,GHz Wi-Fi band is available, thus the bandwidth is restricted to 20\,MHz or 40\,MHz, which limits the accuracy that can be achieved with the \gls{FTM} protocol \cite{noauthor_wi-fi_nodate}. More specifically, according to \cite{noauthor_wi-fi_nodate}, the 90${}^\text{th}$ percentile of the error is expected to be 8\,m for 40\,MHz, 4\,m for 20\,MHz, and 2\,m for 80\,MHz.

\subsection{Measurement Scenarios}

To analyze the performance of the ESP32-S2 module using Wi-Fi \gls{RTT}, two different measurement campaigns were planned. One of them was carried out in an outdoor environment, away from buildings and possible interferences. The second one was carried out in an indoor scenario, in a showroom of the \gls{CITIC}, at the University of A Coruña, Spain. The ESP32-S2-Saola-1 development boards from Espressif Systems were used in both cases. Such modules were mounted on tripods, and then placed in a series of known positions in order to compare the distance estimates with the actual values. Some of the modules were used as Wi-Fi \gls{FTM} responders (acting as beacons) and others as communication initiators (acting as tags).

\subsubsection{Indoor Scenario}
\label{subsec:measurements_scenario_indoor}
\begin{figure}[!t]	
\frame{\includegraphics[width=\columnwidth]{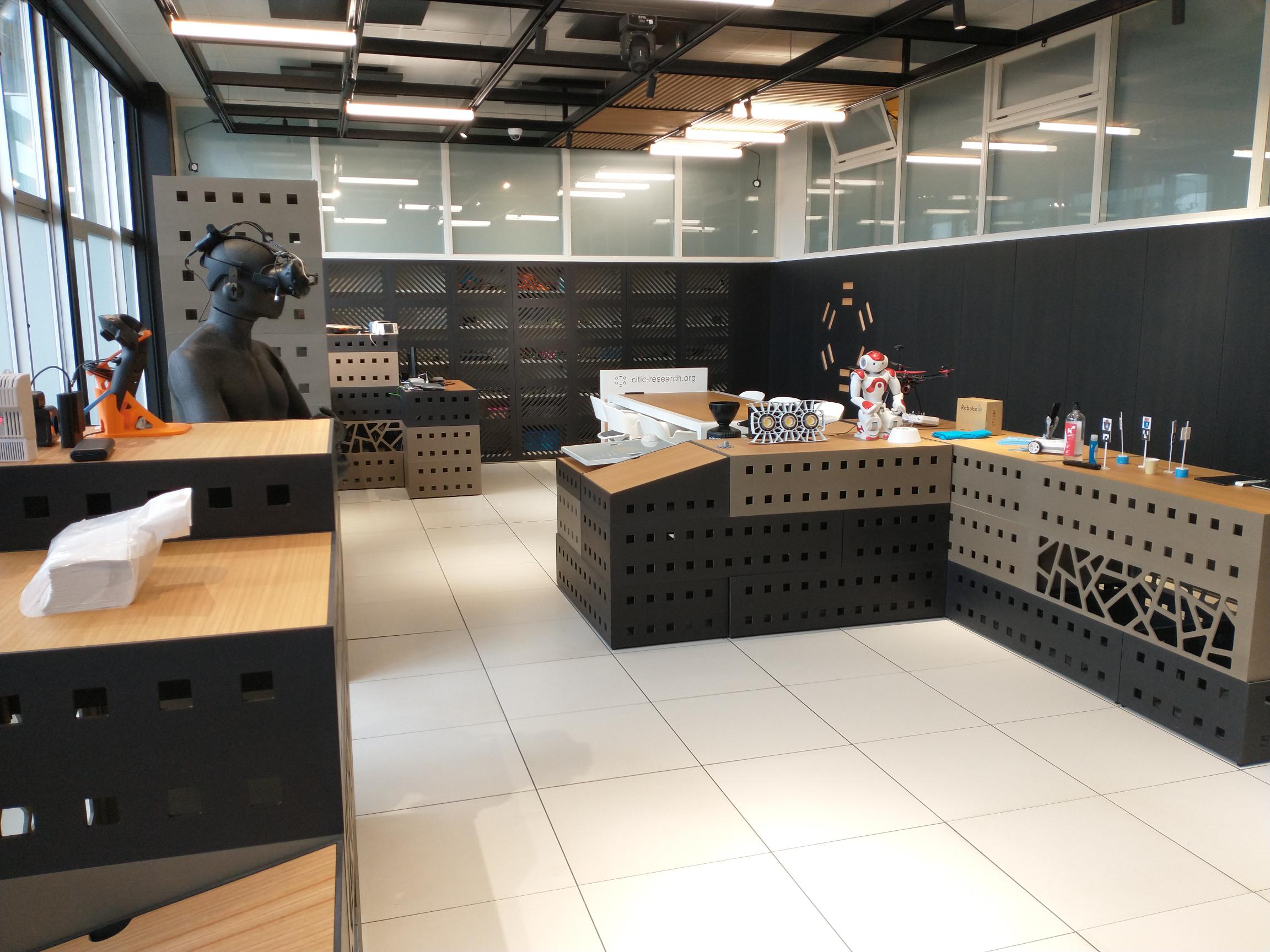}}
\caption{Indoor scenario.\label{fig:measurements_setup_indoor_photo}}
\end{figure}

The \gls{CITIC} showroom (see \cref{fig:measurements_setup_indoor_photo}) at the University of A Coruña, Spain, was selected as an indoor environment. This is a space dedicated to the dissemination of the research activities carried out in the center. It is a room with many obstacles inside and surrounded by sources of interference such as other Wi-Fi networks (more than 7 were detected with a high signal level from inside the room) coexisting with Zigbee sensor networks and Bluetooth devices. The dimensions of the room are 12\,m long and 6\,m wide, with a height of 3.2\,m.

\begin{figure}[!t]	
\centering
\includegraphics[width=\columnwidth]{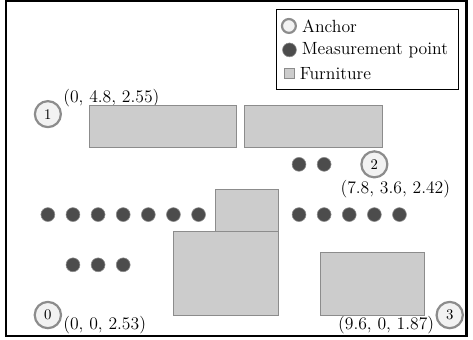}
\caption{Indoor measurement setup.\label{fig:measurements_setup_indoor}}
\end{figure}

To have a greater diversity of measurements, reduce data collection time, and have a dataset applicable to future positioning algorithms, four ESP32-S2 devices were placed at the corners of the room acting as beacons, while the communication initiating tag was moved to different points along a straight line. This tag was placed at a height of 2\,m. A schematic of the anchors and measurement positions is shown in \cref{fig:measurements_setup_indoor}. Due to some of the obstacles present in the room, measurements could not be taken in certain areas. At each point, measurements were taken for 180\,s and the ESP32-S2 devices were configured to work with a 40\,MHz bandwidth, which is the maximum bandwidth allowed at the 2.4\,GHz band. Notice that, due to the relative placement between tag and beacons in this scenario, measurements were obtained at different distances and with different relative angles between transmitter and receiver.

\begin{figure}[!t]	
\frame{\includegraphics[width=\columnwidth]{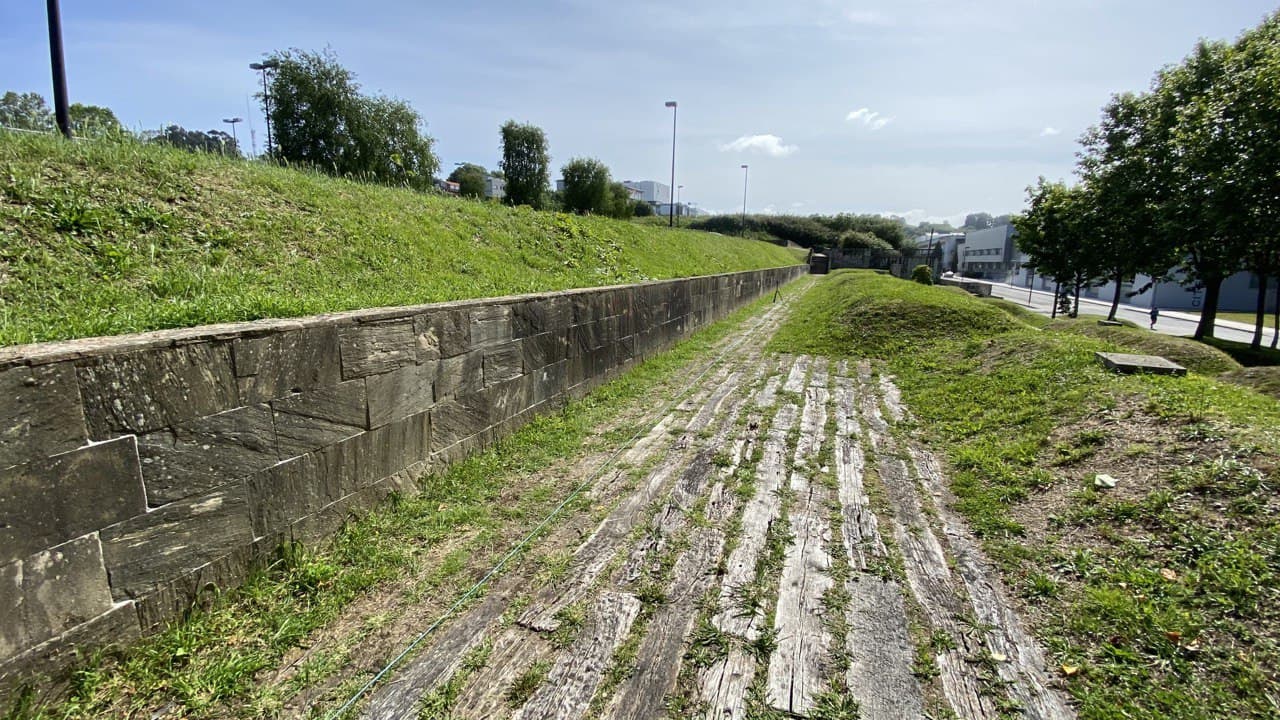}}
\caption{Outdoor scenario.\label{fig:measurements_setup_outdoor}}
\end{figure}  

\subsubsection{Outdoor Scenario}
\label{subsec:measurements_scenario_outdoor}

In order to obtain more information about the capabilities of the ESP32-S2 module for ranging operations with the Wi-Fi \gls{FTM} protocol, another measurement campaign was performed in an outdoor environment. In this way, we tried to minimize the possible effect of interference and multipath typically found in indoor scenarios. In fact, in this scenario, not a single nearby Wi-Fi network was detected that could be a source of interferences. To carry out the measurements, we selected an area of the campus at the University of A Coruña away from the buildings. Only two modules were used in the experiment, one acting as a tag and the other as a beacon. They were mounted on tripods and placed one in front of the other at the same height (1.75\,m). The measurements were taken during 120\,s at distances varying from 1\,m to 20\,m, at 1\,m intervals. Two different measurement campaigns were performed, one with the modules configured with a bandwidth of 20\,MHz, and the other with a bandwidth of 40\,MHz. \cref{fig:measurements_setup_outdoor} shows a picture of the outdoor measurement scenario.

\subsection{Software Development}

In order to collect the measurements, several software modules were implemented. On the one hand, a new firmware was programmed for the ESP32-S2s acting as beacons. In this case, the application configured the module as a Wi-Fi \gls{AP} with the \gls{FTM} responder features enabled. On the other hand, another firmware was programmed for the module acting as a tag. In this case, the functionality of this firmware was divided into three main parts: \textit{1)} a scan of \glspl{AP} capable of responding to an \gls{FTM} request was performed, \textit{2)} once this list was obtained, the module entered in a loop of \gls{FTM} requests with each of the \glspl{AP} detected in the previous phase, and \textit{3)} the measurements obtained were written to the serial port of the development board. 

To assess the collected data from a PC, a \gls{ROS} environment was deployed on a Raspberry Pi 3B. \gls{ROS} \cite{rosweb} is a software commonly used in robotics and allows for decoupling communications between hardware devices (typically sensors) and the software elements in charge of processing their data. Within \gls{ROS}, the fundamental working element is the node. A \gls{ROS} node can publish information in a given topic (a text string similar to a file path or a URL) and can also subscribe to the data available from topics published by other nodes. In our case, we implemented a node capable of collecting the \gls{FTM} measurements captured from the ESP32-S2 (connected via USB port) and publishing them in the \gls{ROS} ecosystem. On the PC we used \emph{rosbag}, an application included by default in \gls{ROS} that allows for the subscription and subsequent storage of measurements in different logs for later analysis. Those measurements were sent from the Raspberry Pi to the PC via an Ethernet link. Note that the choice of \gls{ROS} in this work has been solely to take advantage of its communication system and its measurement collection tools, and not because the nature of what is described here is related to the field of robotics.

As part of the contributions of this paper, the code of all these software modules is publicly available as open source\footnote{Source code can be downloaded from \textit{https://github.com/valentinbarral/*}, where * is one of these project identifiers: \textit{esp32s2-ftm-tag}, \textit{esp32s2-ftm-anchor}, \textit{rosftm}, or \textit{rosmsgs} (see \cite{esp32ftmtag, esp32ftmanchor, rosftmnode, rosftmmsgs}).}.

\begin{figure}[!t]
\includegraphics[width=\columnwidth]{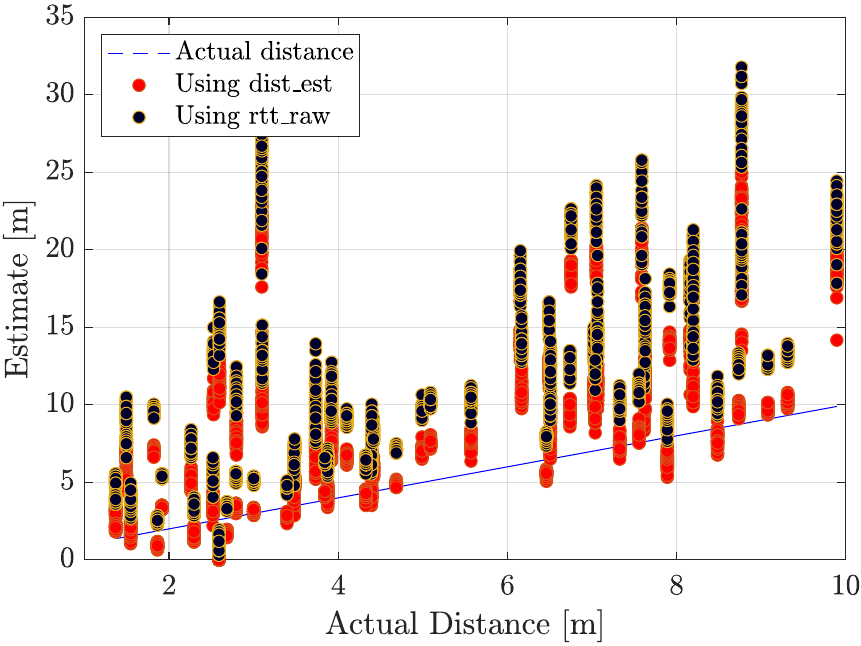}
\caption{Actual versus estimated distance in the indoor scenario.
\label{fig:measurements_analysis_abs_error_indoor}}
\end{figure}  

\section{Measurements Analysis}
\label{sec:measurement_analysis}

This section shows a descriptive analysis of the measurements obtained in the different campaigns. Each measurement includes the following parameters:

\begin{itemize}
    \item {\tt anchorId}: Identifier of the module that acted as beacon in the measurement.
    \item {\tt rtt\_raw}: \gls{RTT} value averaged among the different frames sent and expressed in nanoseconds.
    \item {\tt rtt\_est}: \gls{RTT} estimation provided by the ESP32-S2 firmware in nanoseconds.
    \item {\tt dist\_est}: Distance estimation in meters. Internally, \texttt{rtt\_est} is used to calculate this value.
    \item {\tt num\_frames}: Number of frames successfully sent during the \gls{RTT} communication.
    \item {\tt frames}: A list of all successfully sent frames.
\end{itemize}

Each individual frame includes the following information:

\begin{itemize}
    \item {\tt rssi}: Received signal strength in dBm.
    \item {\tt rtt}: \gls{RTT} value for that frame in nanoseconds.
    \item {\tt t1}: Outgoing timestamp of the first packet from the sender in picoseconds.
    \item {\tt t2}: Timestamp of the reception of the ranging request at the receiver expressed in picoseconds.
    \item {\tt t3}: Timestamp (in picoseconds) of the response message at the receiver.
    \item {\tt t4}: Timestamp (in picoseconds) of the reception of the response message from the receiver at the sender.
\end{itemize}

The results corresponding to indoor and outdoor campaigns are analyzed in \cref{subsec:measurements_analysis_indoor} and \cref{subsec:measurements_analysis_outdoor}, respectively.

\subsection{Analysis of the Indoor Measurements}
\label{subsec:measurements_analysis_indoor}
The first information we extract from the measurements captured in the indoor environment is the distance estimate. The ESP32-S2 provides two \gls{RTT} parameters: {\tt rtt\_raw}, which corresponds to the average of the values obtained in the different frames, and {\tt rtt\_est}, which is estimated by the chip. How the {\tt rtt\_est} estimate is obtained from the raw value is not documented by the manufacturer. These \gls{RTT} parameters can be easily translated into a distance estimate $d$ using the formula
\begin{align}
d=\frac{\texttt{rtt} \cdot c}{2}
\label{eq:time_to_distance}
\end{align}
where {\tt rtt} is the \gls{RTT} value, and $c$ is the speed of light. Note that depending on the value of {\tt rtt} used, {\tt rtt\_est} or {\tt rtt\_raw}, two different distance estimates can be obtained. The ESP32-S2 firmware uses {\tt rtt\_est} to provide an estimate of the distance between the devices available as {\tt dist\_est}.

\begin{figure}[!t]
\includegraphics[width=\columnwidth]{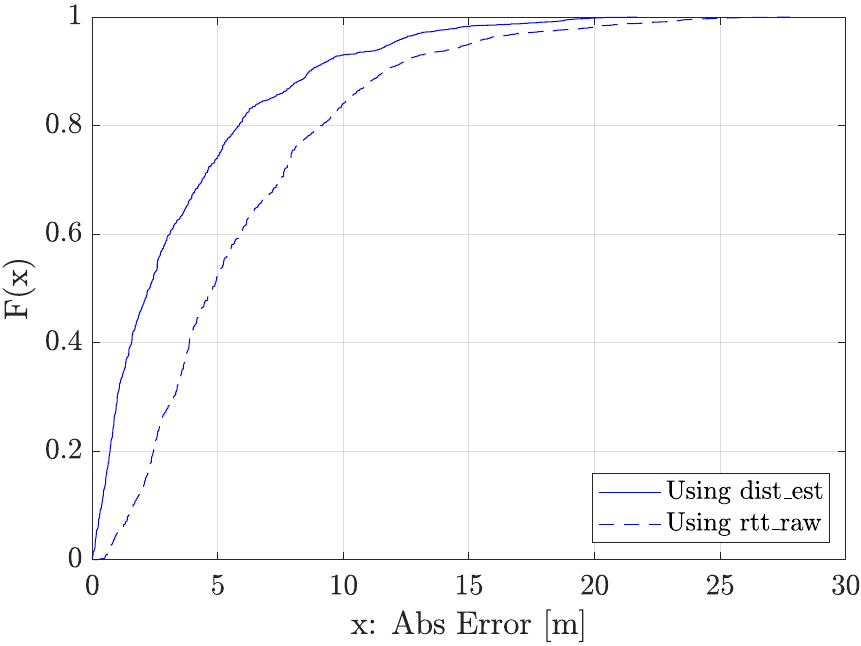}
\caption{ECDF of the absolute error in the indoor scenario.\label{fig:measurements_analysis_ecdf_error_indoor}}
\end{figure} 

\cref{fig:measurements_analysis_abs_error_indoor} shows the captured samples, their actual position, and the two estimates: the one provided by the chip itself, {\tt dist\_est}, and the value obtained from {\tt rtt\_raw} using \cref{eq:time_to_distance}. The values obtained in this indoor environment are clearly very inaccurate. There are large differences between actual and estimated values. It can be seen that the estimates based on the {\tt rtt\_raw} value lead to the worst results, whereas those based on the {\tt dist\_est} achieve a significant improvement in terms of accuracy. The error level obtained can be better seen in \cref{fig:measurements_analysis_ecdf_error_indoor}, where the \gls{ECDF} of the absolute error is shown for both cases. Besides observing the general level of inaccuracy of the measurements, with errors up to 20\,m in some cases, we can see that the distance estimates obtained by the chip are able to improve those from the raw \gls{RTT} samples. As an example, by means of the ESP32-S2 estimation, about 75\,\% of the measurements are below 5\,m of error, whereas using the {\tt rtt\_raw} parameter only 50\,\% are below such a threshold.

\begin{figure}[!t]	
\includegraphics[width=\columnwidth]{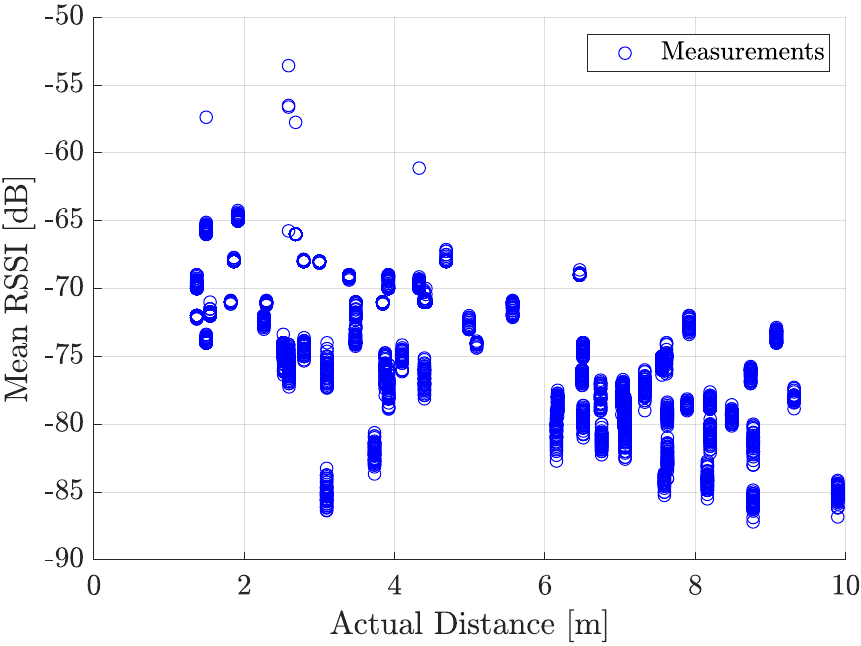}
\caption{Mean RSSI for each measurement versus actual distance for the indoor scenario.\label{fig:measurements_analysis_rss_indoor}}
\end{figure}

Another parameter of interest is the \gls{RSSI} value. \cref{fig:measurements_analysis_rss_indoor} shows the different measurements in the indoor scenario according to their \gls{RSSI}  value and the actual distance at which they were taken. Note that the ESP32-S2 does not generate an \gls{RSSI}  value for each \gls{FTM} sample, but rather for each frame transmitted within an \gls{FTM} communication. In this case, the \gls{RSSI} values shown in \cref{fig:measurements_analysis_rss_indoor} correspond to the average of all the frames in each \gls{FTM} sample. It can be observed that there is a slight decay of the energy as the distance between the devices increases. However, the strong multipath propagation in the indoor environment leads to a severe small-scale fading, and some samples are obtained with high \gls{RSSI} values at long distances and low \gls{RSSI} values at short distances.

\subsection{Analysis of the Outdoor Measurements}
\label{subsec:measurements_analysis_outdoor}

\begin{figure}[!t]
\includegraphics[width=\columnwidth]{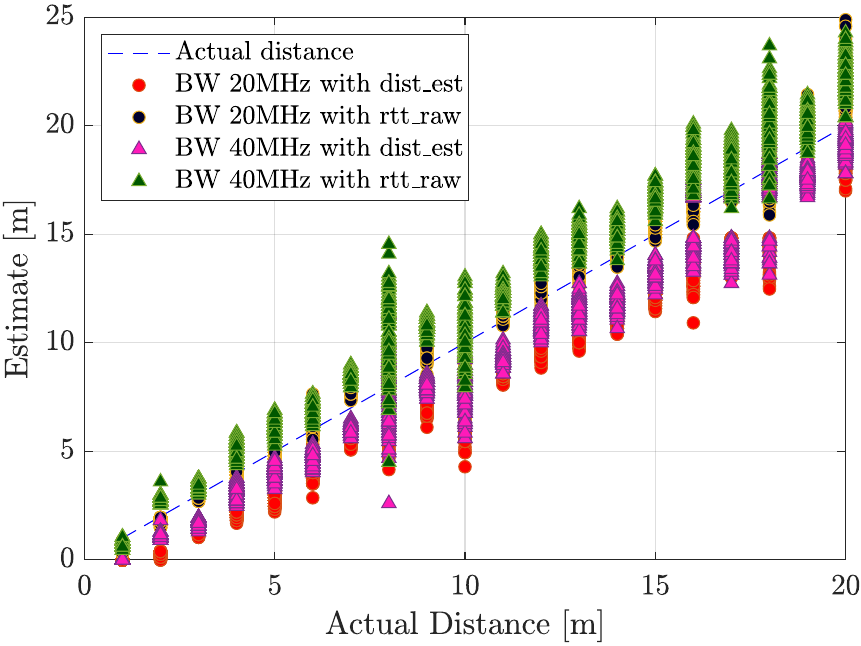}
\caption{Comparison between actual and estimated distances for both outdoor datasets, and for bandwidth values of 20\,MHz and 40\,MHz.\label{fig:measurements_analysis_actual_vs_estimated}}
\end{figure}  

\begin{figure}[!t]	
\includegraphics[width=\columnwidth]{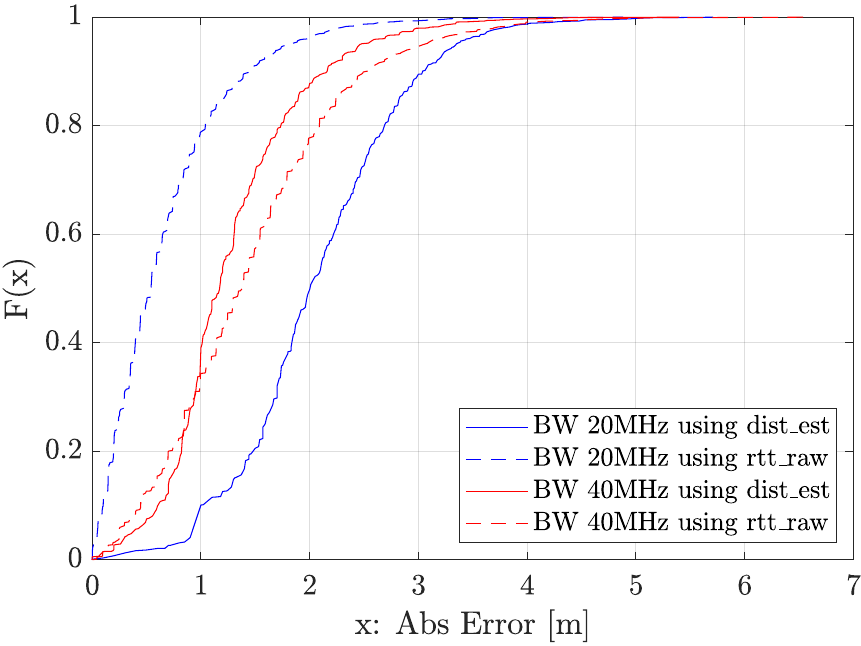}
\caption{ECDF of the absolute error for both outdoor datasets and for both bandwidth values (20\,MHz and 40\,MHz).\label{fig:measurements_analysis_ecdf_20_vs_40}}
\end{figure}  

\cref{fig:measurements_analysis_actual_vs_estimated} shows the distance estimates obtained for the two bandwidth values considered (20\,MHz and 40\,MHz) versus the actual distance. As in the indoor case, both the distance values provided directly by the chip ({\tt dist\_est}) and those generated from the temporal {\tt rtt\_raw} values from \cref{eq:time_to_distance} are shown. In \cref{fig:measurements_analysis_actual_vs_estimated}, it can be seen how the estimated values are much closer to the actual ones than in the indoor case (shown in \cref{fig:measurements_analysis_abs_error_indoor}) for both the 20\,MHz and 40\,MHz bandwidth values. However, in this scenario, a counter-intuitive phenomenon can be seen: the estimates generated with the raw \gls{RTT} values ({\tt rtt\_raw}) are closer to the actual values than the estimates provided by the chip. Such a phenomenon can be clearly observed in \cref{fig:measurements_analysis_ecdf_20_vs_40}, where the estimates from the ESP32-S2 are noticeably worse than those obtained directly from {\tt rtt\_raw}. This effect is clearly visible in the measurements corresponding to the 20\,MHz dataset, whereas in the 40\,MHz datasets both values are more similar. This apparent inconsistency is discussed in more detail in \cref{subsec:est_rtt_vs_raw}. \cref{fig:measurements_analysis_ecdf_20_vs_40} also shows that the raw values from the 20\,MHz configuration ({\tt rtt\_raw}) yield better results than those obtained with 40\,MHz, whereas with the values estimated by the chip ({\tt dist\_est}), this effect is reversed, and the 40\,MHz results are slightly better.

\begin{figure}[!t]	
\includegraphics[width=\columnwidth]{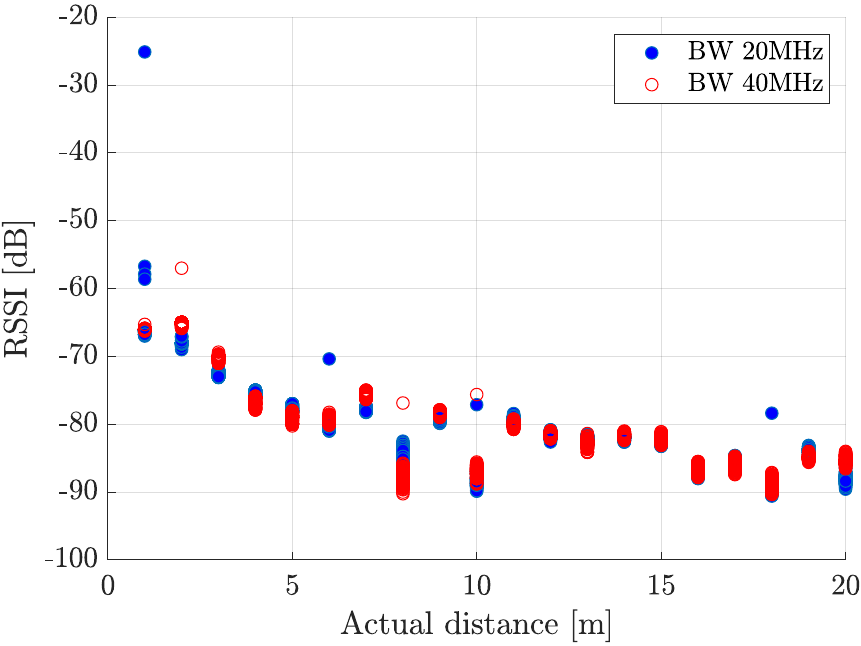}
\caption{RSSI versus actual distance for both outdoor datasets, with bandwidth values of 20\,MHz and 40\,MHz.\label{fig:measurements_analysis_rss_20_vs_40}}
\end{figure}

Finally, \cref{fig:measurements_analysis_rss_20_vs_40} shows the \gls{RSSI} values corresponding to the two bandwidths obtained in the outdoor scenario. We can see that the \gls{RSSI} values exhibit less small-scale fading than in the indoor case (see \cref{fig:measurements_analysis_rss_indoor}), and they are similar regardless of the bandwidth. The energy clearly decreases with distance, although there are points where  multipath effects are observed, probably caused by the signal bouncing off the ground or other obstacles close to the measurement area.

\subsection{Estimated RTT versus Raw RTT}
\label{subsec:est_rtt_vs_raw}

After the analysis of the data captured in the different measurement campaigns, one of the most surprising aspects detected was the difference in accuracy between the distance estimated by the ESP32-S2 and the value computed directly from the raw \gls{RTT} values. While in the indoor case the estimation improved the raw accuracy, this was not the case in the outdoor measurements, specially for the case with a bandwidth of 20\,MHz. In the latter case, the ESP32-S2 estimates were clearly worse than those obtained using the {\tt rtt\_raw} value directly.

\begin{figure}[!t]	
\includegraphics[width=\columnwidth]{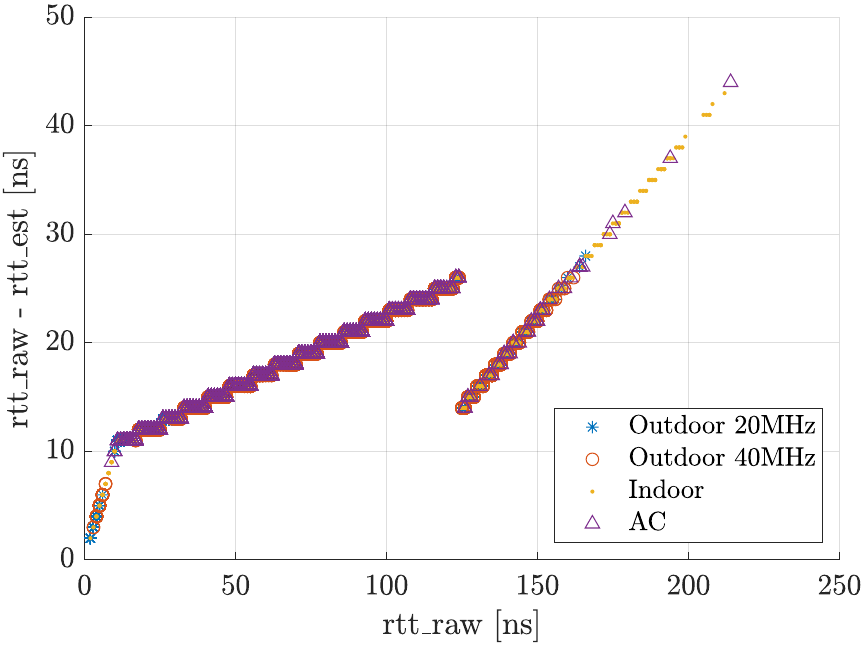}
\caption{$\texttt{rtt\_raw} - \texttt{rtt\_est}$ for all datasets.\label{fig:measurements_rtt_diff}}
\end{figure}

To try to detect the reason for this behavior, and since there is no documentation from the manufacturer about how the distance estimate is generated by the device, a study of the differences between the {\tt rtt\_raw} and {\tt rtt\_est} values was carried out. \cref{fig:measurements_rtt_diff} shows the difference between these two values for all captured measurements (indoors and outdoors) with respect to the {\tt rtt\_raw} value, in which three linear correction models that are directly related to the {\tt rtt\_raw} value itself can be identified. That is, in view of these values, it seems clear that, to generate the {\tt rtt\_est} value, the ESP32-S2 simply applies a correction factor to {\tt rtt\_raw} that depends linearly on its value. Such a linear relationship has a different gradient depending on whether the values are below 10\,ns, between 10\,ns and 124\,ns, or above 124\,ns. Probably, due to the still recent implementation of the \gls{FTM} support on these chips, the chosen correction is not the best. Although it is difficult to say for sure, it seems that an attempt was made to improve the accuracy in indoor environments (a typical scenario to use a Wi-Fi module) at the cost of worsening the results in better conditions, with a good \gls{LOS}, and in absence of obstacles or significant interference between the devices. What does seem clear is that these thresholds, located at 10\,ns and 124\,ns, are consistent across all measurement scenarios, so they must be hard-coded in the ESP32-S2 firmware.

\section{Proposed Estimator Based on Machine Learning}
\label{sec:ml}

After observing the limitations of the current ESP32-S2 estimation algorithm described in \Cref{subsec:est_rtt_vs_raw}, an opportunity appears to propose a better alternative. This new method should be able to operate within the computational limitations of the ESP32-S2, so that it must compute the distance estimate from the {\tt rtt\_raw} in real time and for each \gls{FTM} communication, providing a more accurate result. Having hundreds of data records after the measurement campaigns, one of the direct approaches to the problem would be using \gls{ML} techniques to generate a regression model of the desired correction. This approach is very common to deal with this type of problems \cite{dvorecki2019machine,wymeersch2012machine}, and was also used previously by the authors in previous works for the \gls{UWB} technology \cite{barral_environmental_2019}.

As training features, we selected the {\tt rtt\_raw} values and the average \gls{RSSI} values, discarding other parameters that did not provide any improvement in a preliminary study, such as the number of frames successfully transmitted or the variance of {\tt rtt\_raw} for each sample. To create the training set, we randomly selected 70\,\% of the samples from each of the three datasets (the one corresponding to the indoor measurements and the two outdoors) and merged them into a single one. The remaining samples were reserved as test sets. At the time of training, cross-validation was used with five folds, an adequate number for the not excessively large size of the training set (around 7000 samples). Bayesian optimization \cite{pelikan1999boa} was used to search for the best hyperparameters of each algorithm.

The considered \gls{ML} algorithms were regression trees, \glspl{SVM}, \gls{GP} regression, and shallow neural networks. Their details can be found in \cref{subsec:ml_algorithms}. More complex techniques, such as those based on deep learning, were discarded because of their computational needs, out of the capabilities of the ESP32-S2, and because the number of samples available was too low to apply these techniques with confidence.

\subsection{Description of the ML Algorithms}

\begin{figure}[!t]	
\includegraphics[width=\columnwidth]{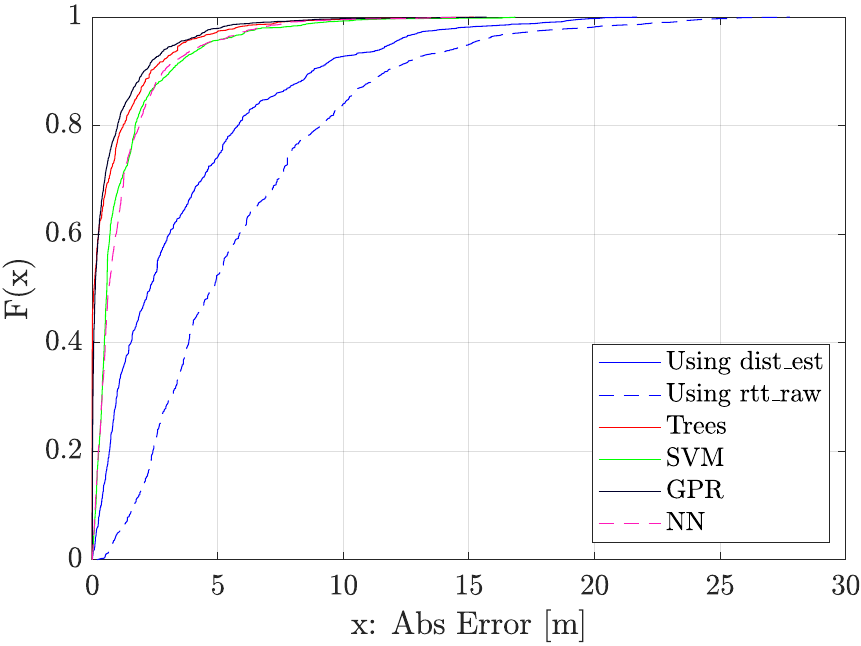}
\caption{ECDF of the absolute error. Test set: indoors with a bandwidth of 20\,MHz.\label{fig:ml_ecdf_indoor}}
\end{figure}  

\label{subsec:ml_algorithms}
The considered \gls{ML} algorithms are described below. All of them are classical algorithms in the literature, and they are widely used in regression problems.
\begin{itemize}
    \item Regression trees: This is an intuitive technique based on binary search trees applied to the regression problem \cite{moisen2008classification}. The idea of these algorithms is to associate homogeneous input sets with the same output. The minimum leaf size is a fundamental parameter in this type of algorithms since, depending on its value, the trees can easily tend to overfitting. In the case presented in this work, the regression tree had a minimum leaf size equal to four. 
    \item \gls{SVM}: This is another classic \gls{ML} algorithm originally described in \cite{vapnik1995nature} that can be used in both classification and regression problems \cite{drucker1997support}. Among the hyperparameters that can be configured in this algorithm, the main one is the type of kernel used to map the input elements in the $n$-dimensional feature space in which the regression process is applied. In the implementation considered in this work, a Gaussian kernel was chosen with the form:
    \begin{equation}
        k\left(\mathbf{x}_{i}, \mathbf{x}_{j}\right)=\exp \left(-\left\|\mathbf{x}_{i}-\mathbf{x}_{j}\right\|^{2}\right)
    \end{equation}
    being $\mathbf{x}_{i},\mathbf{x}_{j} \in R^{n}$ with $n=2$ since the input vectors consider two features in our case.
    \item \gls{GP}: This is a generalization of the Gaussian probability distribution in which the behavior of a function is described. Using this idea, regression and classification models are built with high accuracy and performance \cite{rasmussen2003gaussian}. In the implementation considered in this work, we considered an exponential kernel:
    \begin{equation}
    k\left(\mathbf{x}_{i}, \mathbf{x}_{j} \right)=\sigma_{f}^{2} \exp (-r/\sigma_{l})
    \end{equation}
    being $\mathbf{x}_{i},\mathbf{x}_{j} \in R^{n}$, $\sigma_{f}$ is the signal standard deviation, and $\sigma_{l}$ is the characteristic length scale. Finally, $r$ is defined as the euclidean distance between the vectors $\mathbf{x}_{i}$ and $\mathbf{x}_{j}$:
    \begin{equation*}
        r = \sqrt{\left(\mathbf{x}_{i}-\mathbf{x}_{j}\right)^{T}\left(\mathbf{x}_{i}-\mathbf{x}_{j}\right)}
    \end{equation*}
    After the validation phase, the kernel parameters for the best case were set to $\sigma_{f}=4.6873$ and $\sigma_{l}=0.7051$.
    \item \Gls{NN}: The last of the considered classical algorithms was a fully connected feedforward \gls{NN}. This is one of the most classic \gls{ML} algorithms, already referenced in the middle of the last century, and a pillar of the most complex networks used nowadays in deep learning \cite{schmidhuber2015deep}. In this work, we implemented a simple configuration with a single hidden layer consisting of 100 neurons with a \gls{ReLU} activation function, and a linear activation function in the output layer. All these parameters, as in the rest of the algorithms, were obtained using the Bayesian optimization mechanism limited to 50 iterations.
\end{itemize}
\cref{table:ml_parameters} shows a summary of the parameters of each algorithm.

\begin{figure}[t]	
\includegraphics[width=\columnwidth]{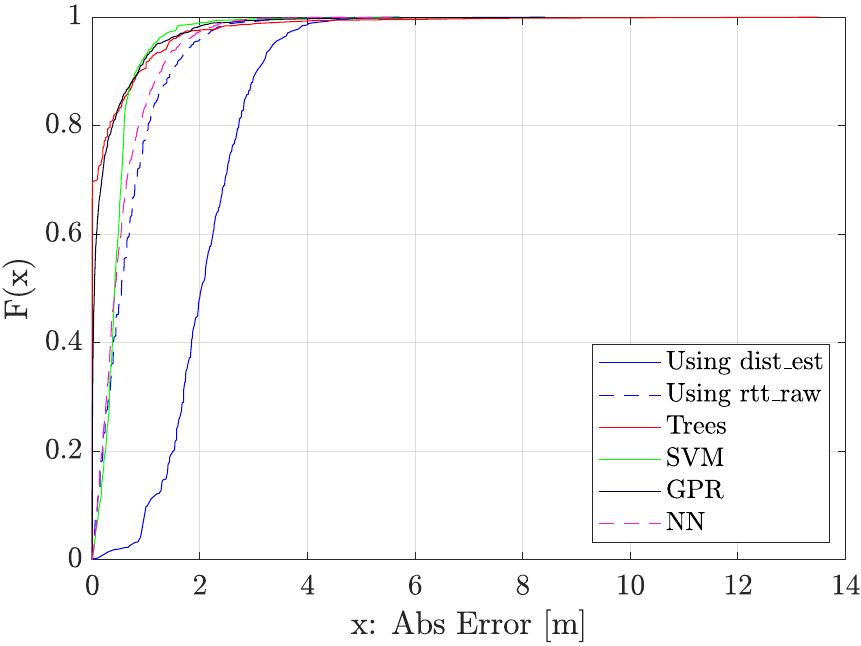}
\caption{ECDF of the absolute error. Test set: outdoors with a bandwidth of 20\,MHz.\label{fig:ml_ecdf_20}}
\end{figure}  

\begin{table}[]
\centering
\caption{Main parameters of tested ML algorithms.}
\label{table:ml_parameters}
\begin{tabular}{ll}
\toprule
Algorithm & Main Parameters\\
\midrule
Regression Trees    & Minimum leaf size = 4\\
\hdashline[1pt/1pt]\noalign{\vskip 0.5ex}
SVM                 & Kernel: Gaussian\\
\hdashline[1pt/1pt]\noalign{\vskip 0.5ex}
\multirow{2}{*}{GP} & $\sigma_{f}=4.6873$\\
                    & $\sigma_{l}=0.7051$\\
\hdashline[1pt/1pt]\noalign{\vskip 0.5ex}
\multirow{4}{*}{NN} & Number of hidden layers = 1\\
                    & Number of neurons = 100\\
                    & Activation function = ReLU\\
                    & Activation function output = linear\\
\bottomrule
\end{tabular}
\end{table}

\subsection{Results of the ML Algorithms}

This section shows the results obtained after applying the trained algorithms to the different test sets, each one coming from a different dataset. \cref{fig:ml_ecdf_indoor} shows the \gls{ECDF} of the absolute error in the indoor scenario in which all the trained ML algorithms exhibit a clear improvement over the estimates from the ESP32-S2. Even with a simple algorithm, such as the regression trees, a noticeable improvement is achieved for most of the samples, being able to place 80\,\% of them below 1.5\,m error, whereas with the original estimates, the corresponding error increases up to 6\,m.

\cref{fig:ml_ecdf_20} shows the \gls{ECDF} of the absolute error in the outdoor scenario considering a bandwidth of 20\,MHz. This was the case in which the ESP32-S2 estimates were clearly worse than those computed using the {\tt rtt\_raw} parameter directly. In view of \cref{fig:ml_ecdf_20}, all the ML algorithms outperform both the ESP32-S2 estimates and the values computed from {\tt rtt\_raw}.

\begin{figure}[!t]	
\includegraphics[width=\columnwidth]{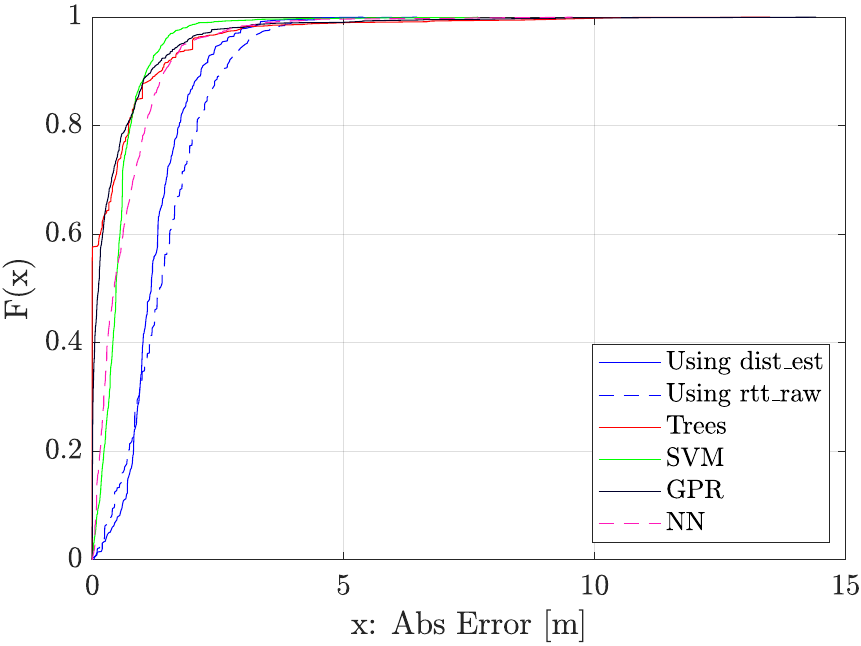}
\caption{ECDF of the absolute error. Test set: outdoors with a bandwidth of 40\,MHz.\label{fig:ml_ecdf_40}}
\end{figure}

Finally, \cref{fig:ml_ecdf_40} shows the \gls{ECDF} of the absolute error also in the outdoor scenario, but in this case using the 40\,MHz bandwidth configuration. With this configuration, and according to the \gls{ECDF} in \cref{fig:measurements_analysis_ecdf_20_vs_40}, there was originally very little difference between the ESP32-S2 estimates and the values computed from the {\tt rtt\_raw} parameter (although the estimates yielded the best result, as shown in \cref{fig:measurements_analysis_ecdf_20_vs_40}). However, in view of \cref{fig:ml_ecdf_40}, a significant improvement is observed when the \gls{ML} algorithms are employed.

As a conclusion, in view of the results shown in \cref{fig:ml_ecdf_indoor,fig:ml_ecdf_20,fig:ml_ecdf_40}, ML-based estimators improve the accuracy of the measurements to some degree. However, it should be noted that, although the test set did not include any samples in common with the training set, it is true that the general distribution is similar because they all come from the same dataset. For this reason and to assess the robustness of the models, a new experiment was carried out in a completely new measurement scenario. The details and results obtained from this experiment are shown in \cref{sec:test}.

\section{Assessment of the ML Trained Models in a Different Environment}
\label{sec:test}

A proposal was presented in \cref{sec:ml} to use classical \gls{ML} algorithms to estimate the actual distance from the {\tt rtt\_raw} value and the signal level. These algorithms were trained with part of the samples collected in the two proposed scenarios, the outdoor and the indoor one. The results of this training were tested with the the samples from these datasets that had not been included in the training set. In a further step, to validate this approach and check if it can be used in other different environments, a new measurement campaign was carried out in a different scenario. This new environment is described in \cref{subsec:test_scenario}.

\begin{figure}[!t]	
\frame{\includegraphics[width=\columnwidth]{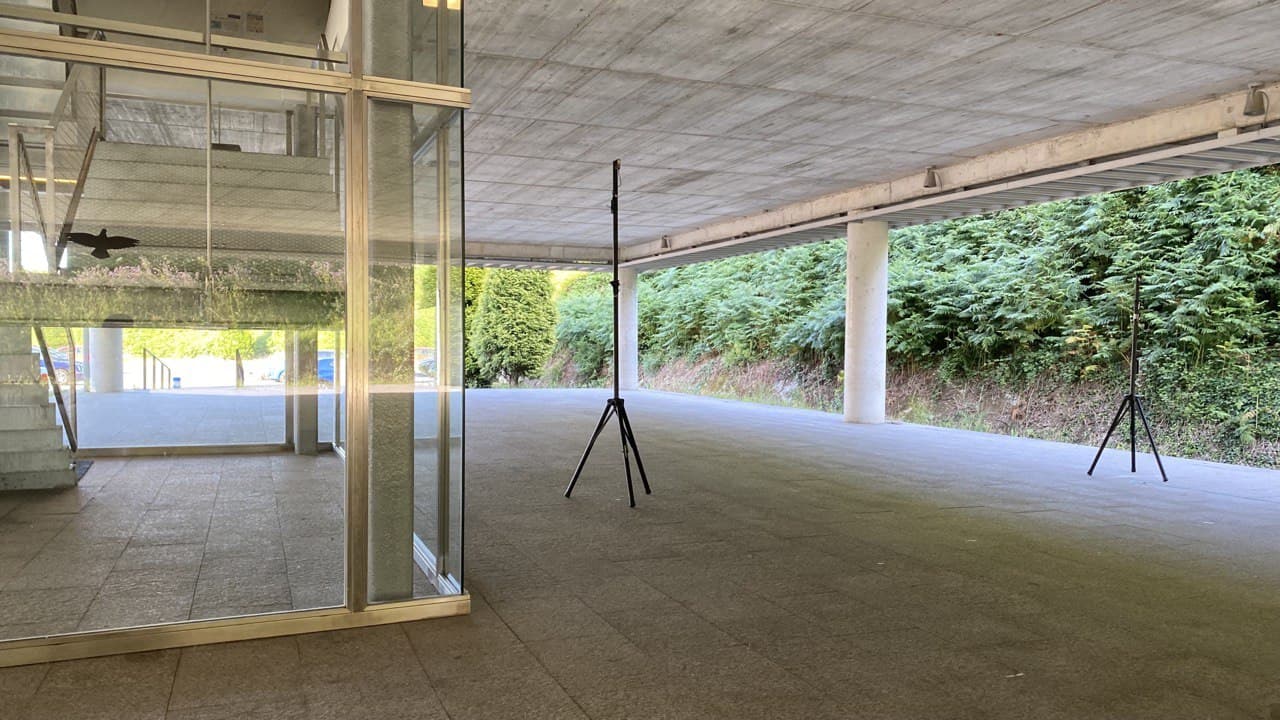}}
\caption{Scenario in the Scientific Area building.\label{fig:test_photo_ac}}
\end{figure}  

\begin{figure}[!t]	
\includegraphics[width=\columnwidth]{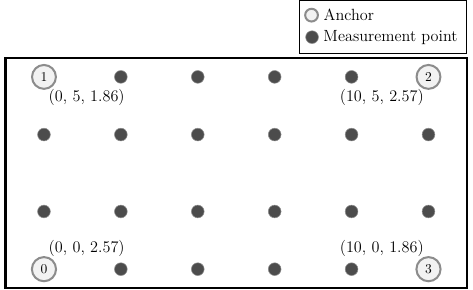}
\caption{Measurement setup of the scenario in the Scientific Area building. Anchors 0 and 1 use a bandwidth of 20\,MHz, whereas anchors 2 and 3 employ 40\,MHz.\label{fig:test_setup_ac}}
\end{figure}  

\subsection{Test Scenario}
\label{subsec:test_scenario}
To test the trained models, we looked for a scenario different from the outdoor and indoor scenarios already considered. The place chosen (see \cref{fig:test_photo_ac}) was the ground floor of the Scientific Area building, a research building located in the Campus of Elviña at the University of A Coruña, Spain. From the point of view of radio propagation, it is a complicated scenario due to its low ceiling, glazed areas, and metal structures located at different points. In addition, being a research building, there are several nearby devices radiating in the 2.4\,GHz band that could affect to some degree the communications with the ESP32-S2. Specifically, after scanning inside the area, 5 to 7 access points with a medium-low signal level were detected.

To perform the measurements, four tripods were placed at the corners of a $10 \times 5$\,m rectangle, mounting an ESP32-S2, configured as an anchor, on each of them (at different heights), as shown in \cref{fig:test_setup_ac}. Anchors labeled with 0 and 1 employed a bandwidth of 20\,MHz, whereas anchors 2 and 3 used 40\,MHz. A series of points were marked within a grid as shown in \cref{fig:test_setup_ac} and measurements were taken with a fifth ESP32-S2 device mounted on another tripod and moved over each of the measurement points. \gls{FTM} measurements were captured for 120\,s at each point using the same software and hardware configuration as in the previous datasets. That is, a Raspberry Pi 3B connected to the ESP32-S2, which acted as a tag, was in charge of reading the measured values through the serial port and publishing them within a \gls{ROS} environment to be later saved in a remote PC. All the samples collected, as well as those from the other datasets, are publicly available to the research community in \cite{2pv8-ze59-21}.

\subsection{Results}
\label{subsec:test_results}

\begin{figure}[!t]	
\includegraphics[width=\columnwidth]{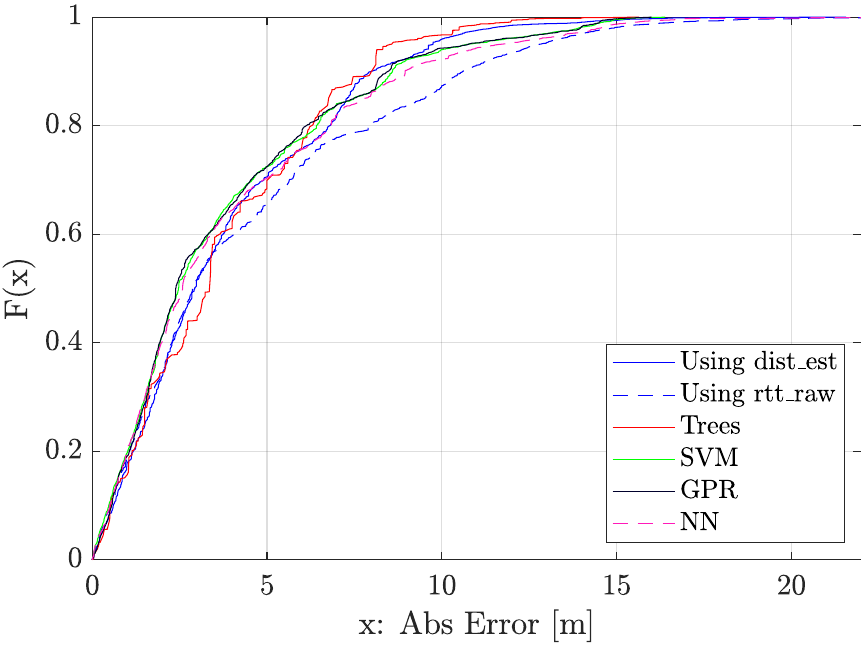}
\caption{ECDF of the absolute error. Test set: Scientific Area building with bandwidth values of 20\,MHz and 40\,MHz.\label{fig:test_results_all}}
\end{figure}

After performing the measurement campaign in the Scientific Area building, the algorithms trained with the other datasets were run on the new samples. All the results from this measurement scenario consider simultaneously values corresponding to 20\,MHz and 40\,MHz bandwidth values. \cref{fig:test_results_all} shows the \gls{ECDF} of the absolute error and several details of interest can be observed. First, we see that the error is slightly lower than in the indoor scenario (see \cref{fig:ml_ecdf_indoor}), but much higher than in the outdoor scenario (see \cref{fig:ml_ecdf_20,fig:ml_ecdf_40}). On the other hand, we see that the ESP32-S2 estimation improves the value computed directly from {\tt rtt\_raw}, although the differences are smaller than in the indoor scenario. These differences are found in the values with the highest level of error, whereas the number of values with errors below 4\,m is practically the same using both approaches.

\begin{figure}[!t]	
\includegraphics[width=\columnwidth]{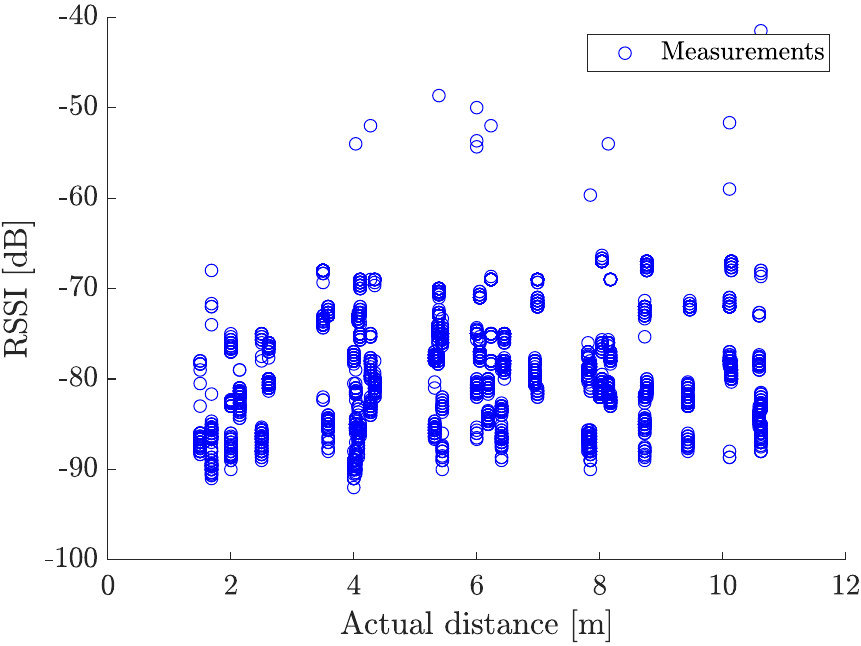}
\caption{Mean \gls{RSSI} on the measurements captured in the Scientific Area building with bandwidth values of 20\,MHz and 40\,MHz. \label{fig:test_results_rss}}
\end{figure}

As for the predictions of the \gls{ML} algorithms, we see in \cref{fig:test_results_all} that we do not achieve a performance improvement similar to that obtained with the other datasets. The values obtained are very close to the ESP32-S2 estimates, slightly better in some cases. The reason for this difference with respect to the results is due to the nature of the chosen scenario and its propagation problems, as well as by the ESP32-S2 antenna and its radiation pattern. \cref{fig:test_results_rss} shows the \gls{RSSI} level distribution of the samples with respect to the distance. There are hardly any differences between the values obtained at a distance of 1.5\,m with respect to those captured at distances of more than 10\,m. If we compare these values with those obtained from the other scenarios (see \cref{fig:measurements_analysis_rss_indoor,fig:measurements_analysis_rss_20_vs_40}), we can see that, in this case, the values are strongly affected by multipath, and there is not a clear energy drop as the distance between the emitter and the receiver increases. This explains, at least in part, why the estimation algorithms in this scenario, which were trained with the {\tt rtt\_raw} features and the \gls{RSSI}, do not perform as well as in the other scenarios. However, it must be said that the estimations are never worse than those provided by the ESP32-S2, hence it is realistic to think that, in less aggressive scenarios, the results will be better and more similar to those obtained in the initial test scenarios. Therefore, \gls{ML} algorithms provide stability and robustness against any kind of scenario, being a more adequate solution than the one included with the ESP32-S2 firmware.

\section{Deployment in the ESP32-S2}
\label{sec:test_deployment}

Taking advantage of the ESP32-S2 capabilities, one of the previously trained algorithms was implemented inside the chip. In particular, due to its simplicity, we implemented the regression trees algorithm. This algorithm, once trained, has very low computational requirements to generate a new estimate. In this way, real-time estimates could be obtained at the same time as the ESP32-S2's own estimates were calculated. Thus, the measurements obtained during the measurement campaign in the Scientific Area building already include an additional {\tt own\_est} parameter that corresponds to the real-time estimation provided by this algorithm.

Hence, the ESP32-S2 firmware was modified to include a new function whose workflow was:

\begin{itemize}
    \item Get the {\tt rtt\_raw} value of the last \gls{FTM} measurement.
    \item Generate the average \gls{RSSI}  of all frames correctly sent within the measurement.
    \item Normalize these values with the same normalization parameters used during the training phase.
    \item Run the regression tree algorithm to estimate a distance value.
    \item Add this new value as the {\tt own\_est} parameter within the measurements written to the serial port.
\end{itemize}

For the implementation, we started from the code generated using the Matlab Coder utility. Once this base was obtained, it was modified and adapted to the needs of the ESP32-S2, both in terms of memory management and use of libraries. This implementation is also publicly available as open source \cite{esp32ftmtag}. The final size of the executable is only 403\,KB, out of a total of 4\,MB available on the ESP32-S2.

\begin{figure}[!t]
\includegraphics[width=\columnwidth]{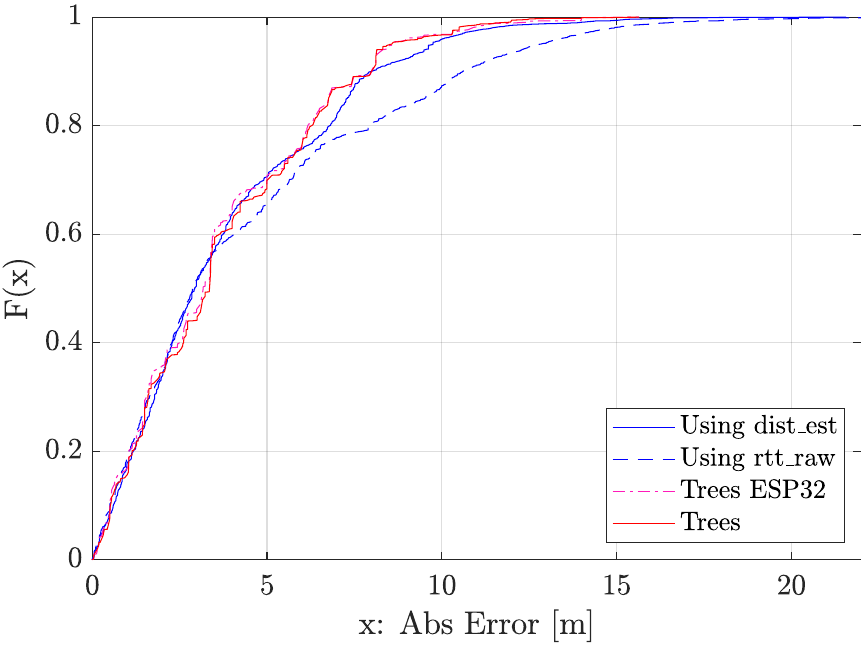}
\caption{ECDF of the absolute error. Test set: Scientific Area building. Offline versus real-time ESP32-S2 implementation considering bandwidth values of 20\,MHz and 40\,MHz. \label{fig:test_results_implementation}}
\end{figure}  

\cref{fig:test_results_implementation} shows the \gls{ECDF} of the absolute value of the two versions: the offline algorithm and the real-time algorithm implemented within the chip. It can be seen how the results are practically identical, the small differences being due to possible inaccuracies in some values at the time of averaging the \gls{RSSI}  or the normalization process. 

\subsection{ESP32 Current Consumption}
\label{subsec:esp32_power}

\begin{figure}[!t]
\frame{\includegraphics[width=\columnwidth]{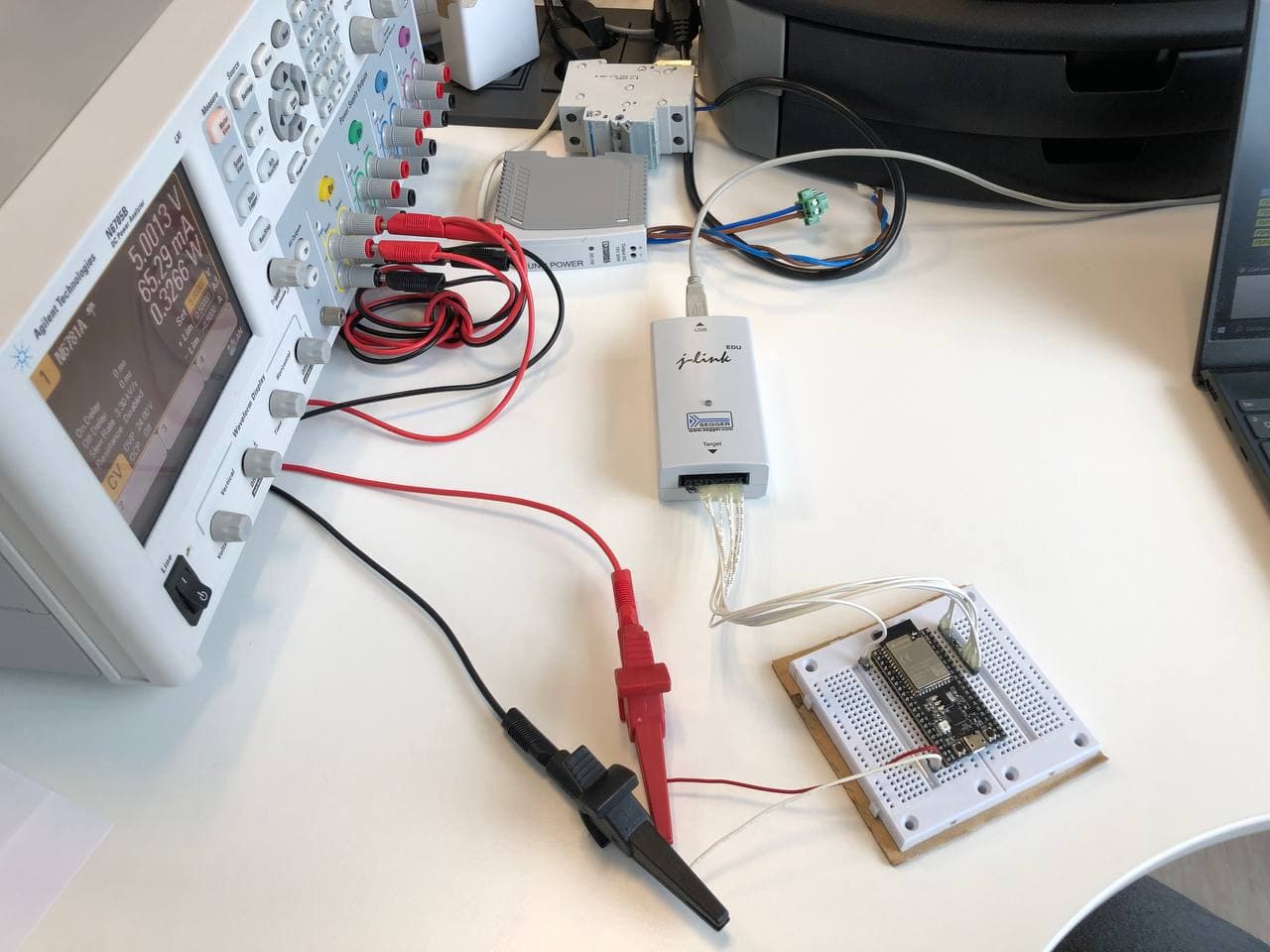}}
\caption{ESP32-S2 current consumption measurement setup. \label{fig:consumption_ftm_setup}}
\end{figure}

Within the IoT world, sensor energy consumption is often a factor of utmost importance. This section shows the current consumption details of the ESP32-S2 when performing positioning tasks using FTM. These measurements were obtained using an N6705 Keysight DC Power Analyzer (see \cref{fig:consumption_ftm_setup}). Notice that the voltage applied to the power supply was set to 5\,V, hence the conversion from current to power consumption is direct.

In order to get the current consumption values, the board was switched to the deep-sleep mode, and the red LED was removed to save energy and to obtain more realistic current measurements. The current consumption obtained was $560.6\,\upmu$A, which basically includes the consumption of the ESP32-S2, the USB-UART bridge, and the \gls{LDO}.
To break down the consumption of each of these components, the ESP32-S2 and the USB-UART bridge were disassembled, and the consumption was measured again. In this way, the \gls{LDO} consumption was obtained, which was $337.9\,\upmu$A. Taking into account the USB-UART bridge, and the ESP32-S2 datasheets \cite{espresssif_systems_esp32-s2_2021}, \cite{CP2102NGQFN28USBXpressUSB2021}, the typical current consumption, when they are not in use, ranges between $195$ and $200\,\upmu$A in the case of the USB-UART bridge, and between $20$ and $25\,\upmu$A in the case of the ESP32-S2. The difference between $560.6\,\upmu$A and $337.9\,\upmu$A results in $222.7\,\upmu$A, which corresponds to the current consumption specified for the ESP32-S2 and the USB-UART bridge. Once these calculations were available, the current consumption of the ESP32-S2 module was estimated, which varies between $222.7 {-} 200 {=} 22.7\,\upmu$A and $222.7 {-} 195 {=} 27.7\,\upmu$A. 

\begin{figure}[!t]
\includegraphics[width=\columnwidth]{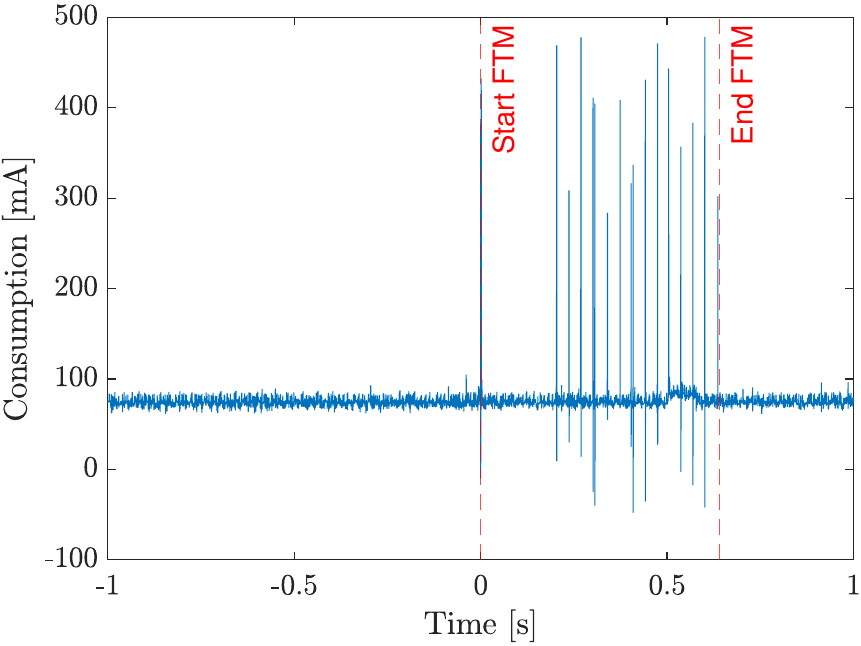}
\caption{ESP32-S2 current consumption while FTM protocol. \label{fig:consumption_ftm}}
\end{figure}  

Finally, after assembling the components that were removed from the board, the current consumption was measured while performing \gls{FTM} with a resolution of $128000$\,points/s. The setup can be seen in \cref{fig:consumption_ftm_setup}, whereas \cref{fig:consumption_ftm} shows the obtained results. When the ESP32-S2 is transmitting, high consumption peaks appear with a maximum of $454$\,mA, and when it is not transmitting, the consumption is almost constant around $73$\,mA. Since an \gls{FTM} operation takes $636$\,ms to complete, during that time the average consumption is $75.6$\,mA. This is a large current value since the \gls{MCU} cannot be switched to the deep-sleep mode when it is transmitting data. If the chip were switched to the deep-sleep mode when it is not transmitting, the average consumption could be reduced significantly. 

\begin{figure}[ht]
\begin{subfigure}{\columnwidth}
\centering
\includegraphics[width=0.7\columnwidth]{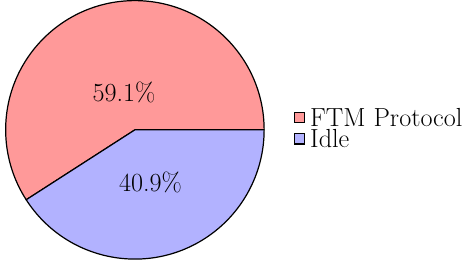}
\caption{Performing a single \gls{FTM} transmission per minute.}
\label{fig_energy_1m}
\end{subfigure}\vspace*{\fill}
\begin{subfigure}{\columnwidth}
\centering
\includegraphics[width=0.7\columnwidth]{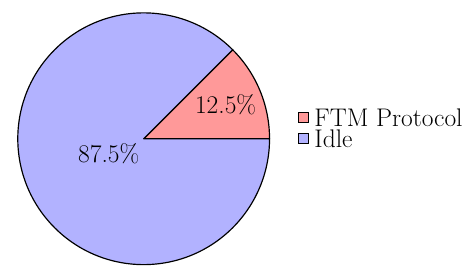}
\caption{Performing a single \gls{FTM} transmission per 10 minutes.}
\label{fig_energy_10m}
\end{subfigure}
\caption{Energy budget for one day performing a different number of \gls{FTM} transmissions.} 
\label{fig_energy_budget}
\end{figure}

This large difference in power consumption between the idle mode and the \gls{FTM} transmission mode can be seen more clearly using the concept of energy budget. This concept describes the energy consumption of the chip depending on the different working modes in which it is configured throughout the day. This idea is similar to the one presented in other works such as \cite{pincheiraCosteffectiveIoTDevices2021,bougueraEnergyConsumptionModel2018}. Thus, we define the daily energy budget as:
\begin{equation}
        E_\text{(daily budget)}= E_\text{(idle)} + E_\text{(FTM)}
\end{equation}
where $E_\text{(daily budget)}$ is the energy consumed by the chip in one day, $E_\text{(idle)}$ is the energy consumed when the chip is in idle mode, and $E_\text{(FTM)}$ is the energy consumed when \gls{FTM} protocol messages are being transmitted. Obviously, this energy consumption will be different depending on the number of \gls{FTM} measurements taken during the day. Thus, as an example, \cref{fig_energy_budget} shows the daily energy budget considering two different intervals. In the case of \cref{fig_energy_1m}, the energy values are considered when performing a single \gls{FTM} measurement per minute, while in \cref{fig_energy_10m} the values correspond to a measurement periodicity of 10 minutes. It can be seen how the impact of the \gls{FTM} measures is very large in the total, even though the chip spends most of the day in idle mode ($98.8$\% of the time when the \gls{FTM} periodicity is 1 minute, $99.89$\% of the time when the \gls{FTM} periodicity is 10 minutes).

In order to verify the impact on the current consumption caused by the proposed solution based on ML, new measurements were taken with and without this functionality activated. \cref{table:ftm_consumption_estimation} shows an estimation of the average current consumption if the number of \gls{FTM} frames were 1 every 10 seconds, 1 every minute, etc, considering a 2000 mAh battery to estimate the battery lifetime. The estimation was made taking into account that the chip was in the deep-sleep mode when it was not transmitting. As it can be seen in \cref{table:ftm_consumption_estimation}, the proposal hardly has an impact on the current consumption.

\begin{table}[]
\centering
\caption{\gls{FTM} current consumption estimation and estimated battery lifetime with a $2000$\,mAh battery.}
\label{table:ftm_consumption_estimation}
\begin{tabular}{m{1.5cm}m{1.9cm}m{1.7cm}m{1.5cm}}
\toprule
Algorithm & FTM measurement period & Current consumption & Estimated battery lifetime \\ \midrule
 \multirow{5}{=}{FTM regression trees} & 10\,s & 5.33\,mA & 15 days \\
 & 1\,min  & 1.36\,mA & 61 days \\
 & 10\,min & 0.64\,mA & 130 days \\
 & 30\,min & 0.59\,mA & 142 days \\
 & 1\,h    & 0.57\,mA & 145 days \\
\hdashline[1pt/1pt]\noalign{\vskip 0.5ex}
 \multirow{5}{=}{FTM Espressif} & 10\,s & 5.29\,mA & 15 days \\
 & 1\,min  & 1.35\,mA & 61 days \\
 & 10\,min & 0.64\,mA & 130 days \\
 & 30\,min & 0.59\,mA & 142 days \\
 & 1\,h    & 0.57\,mA & 145 days \\
\bottomrule
\end{tabular}
\end{table}

\section{Conclusions and Future Work}
\label{sec:conclusions}

In this work we have analyzed the performance of the ESP32-S2 module as a device capable of providing distance estimation using FTM Wi-Fi technology implemented according to the IEEE 802.11-2016 standard. For this purpose, we have performed several measurement campaigns in different scenarios, using several ESP32-S2 devices with a specific firmware. Subsequently, we have evaluated the estimates obtained, resulting in high error values in indoor environments (up to 5\,m for the 75\,\% of the measurements with a bandwidth of 20\,MHz) and lower errors in outdoor environments (up to around 1.5\,m for the 75\,\% of the measurements with a bandwidth of 40\,MHz and up to 2.5\,m when the bandwidth is 20\,MHz). This difference in the error levels is due to several factors. One of them is the utilization of an algorithm based on an energy threshold to perform the temporal marking of the packets during the \gls{FTM} protocol. In this case, multipath and interference problems in indoor environments cause that the first path of the signal be missed, and a bounce is taken instead. In this case, the time of flight of the signal becomes longer than the actual one, and therefore the distance estimation is also longer. 

During the study, it has also been found that the ESP32-S2 employs \gls{RTT} distance estimates different than the ones observed in the frames during \gls{FTM} communications. Although this is not detailed in the ESP32-S2 documentation, we have experimentally verified that a linear correction is applied to the raw \gls{RTT} values depending only on the values themselves. We have found that this approach improves the accuracy in some cases but produces large errors in other cases. 

We have proposed to use classical \gls{ML} algorithms to perform the final distance estimation, taking as training features the raw \gls{RTT} values and the signal level indicator of each sample. We have trained and tested the algorithms, showing an improvement in the absolute error obtained. Additionally, we have implemented one of these trained algorithms inside the ESP32-S2, so that we have been able to generate the estimates in real time and simultaneously with those provided by the chip. An additional measurement campaign was carried out, in a totally different scenario from those considered to obtain the training samples, with the objective of validating the \gls{ML} algorithms and the real-time implementation. The results obtained, despite the difficulties of the environment, demonstrated that our proposal based on \gls{ML} is suitable to address the problem of generating the final estimates from the \gls{RTT} values.

Overall, the \gls{FTM} implementation in the ESP32-S2 is still far from being usable in complex indoor environments. While its performance in outdoor environments exhibited good results, with almost 90\,\% of the samples below 2\,m of absolute error for a bandwidth value of 40\,MHz, in indoor environments only 40\,\% of the samples fall below such a threshold (for a bandwidth value of 20\,MHz). In addition, in the indoor scenario there were very large estimation errors for some of the samples, with almost 10\,\% of the measurements having more than 8\,m of error. Thus, with these results, it would be very difficult to generate 2D or 3D position estimates with an error smaller than the size of a medium-sized room. 

In future works, new experiments will be designed to assess the accuracy obtained in a 3D environment, considering not only the location in a plane, but also the height.

\bibliography{IEEEabrv,main} 

\begin{thebibliography}{10}
\providecommand{\url}[1]{#1}
\csname url@samestyle\endcsname
\providecommand{\newblock}{\relax}
\providecommand{\bibinfo}[2]{#2}
\providecommand{\BIBentrySTDinterwordspacing}{\spaceskip=0pt\relax}
\providecommand{\BIBentryALTinterwordstretchfactor}{4}
\providecommand{\BIBentryALTinterwordspacing}{\spaceskip=\fontdimen2\font plus
\BIBentryALTinterwordstretchfactor\fontdimen3\font minus
  \fontdimen4\font\relax}
\providecommand{\BIBforeignlanguage}[2]{{%
\expandafter\ifx\csname l@#1\endcsname\relax
\typeout{** WARNING: IEEEtran.bst: No hyphenation pattern has been}%
\typeout{** loaded for the language `#1'. Using the pattern for}%
\typeout{** the default language instead.}%
\else
\language=\csname l@#1\endcsname
\fi
#2}}
\providecommand{\BIBdecl}{\relax}
\BIBdecl

\bibitem{khelifiSurveyLocalizationSystems2019}
F.~Khelifi, A.~Bradai, A.~Benslimane, P.~Rawat, and M.~Atri, ``A {Survey} of
  {Localization} {Systems} in {Internet} of {Things},'' \emph{Mobile Networks
  and Applications}, vol.~24, Jun. 2019.

\bibitem{tedeschiIoTraceFlexibleEfficient2021}
P.~Tedeschi, S.~Bakiras, and R.~Di~Pietro, ``{IoTrace}: {A} {Flexible},
  {Efficient}, and {Privacy}-{Preserving} {IoT}-{Enabled} {Architecture} for
  {Contact} {Tracing},'' \emph{IEEE Communications Magazine}, vol.~59, no.~6,
  pp. 82--88, Jun. 2021, conference Name: IEEE Communications Magazine.

\bibitem{liLocationEnabledIoTLEIoT2021}
Y.~Li, Y.~Zhuang, X.~Hu, Z.~Gao, J.~Hu, L.~Chen, Z.~He, L.~Pei, K.~Chen,
  M.~Wang, X.~Niu, R.~Chen, J.~Thompson, F.~M. Ghannouchi, and N.~El-Sheimy,
  ``Toward {Location}-{Enabled} {IoT} ({LE}-{IoT}): {IoT} {Positioning}
  {Techniques}, {Error} {Sources}, and {Error} {Mitigation},'' \emph{IEEE
  Internet of Things Journal}, vol.~8, no.~6, pp. 4035--4062, Mar. 2021,
  conference Name: IEEE Internet of Things Journal.

\bibitem{brena_evolution_2017}
R.~F. Brena, J.~P. García-Vázquez, C.~E. Galván-Tejada, D.~Muñoz-Rodriguez,
  C.~Vargas-Rosales, and J.~Fangmeyer, ``Evolution of indoor positioning
  technologies: {A} survey,'' \emph{Journal of Sensors}, vol. 2017, 2017,
  publisher: Hindawi.

\bibitem{sakpere_state---art_2017}
W.~Sakpere, M.~Adeyeye-Oshin, and N.~B. Mlitwa, ``A state-of-the-art survey of
  indoor positioning and navigation systems and technologies,'' \emph{South
  African Computer Journal}, vol.~29, no.~3, pp. 145--197, 2017, publisher:
  South African Computer Journal.

\bibitem{laoudias_survey_2018}
C.~Laoudias, A.~Moreira, S.~Kim, S.~Lee, L.~Wirola, and C.~Fischione, ``A
  survey of enabling technologies for network localization, tracking, and
  navigation,'' \emph{IEEE Communications Surveys \& Tutorials}, vol.~20,
  no.~4, pp. 3607--3644, 2018, publisher: IEEE.

\bibitem{seco_smartphone-based_2018}
F.~Seco and A.~R. Jiménez, ``Smartphone-based cooperative indoor localization
  with {RFID} technology,'' \emph{Sensors}, vol.~18, no.~1, p. 266, 2018,
  publisher: Multidisciplinary Digital Publishing Institute.

\bibitem{yang_wifi-based_2015}
C.~Yang and H.-R. Shao, ``{WiFi}-based indoor positioning,'' \emph{IEEE
  Communications Magazine}, vol.~53, no.~3, pp. 150--157, 2015, publisher:
  IEEE.

\bibitem{teran_iot-based_2017}
M.~Terán, J.~Aranda, H.~Carrillo, D.~Mendez, and C.~Parra, ``{IoT}-based
  system for indoor location using bluetooth low energy,'' in \emph{2017 {IEEE}
  colombian conference on communications and computing ({COLCOM})}, 2017, pp.
  1--6.

\bibitem{huang_zigbee-based_2011}
C.-N. Huang and C.-T. Chan, ``{ZigBee}-based indoor location system by
  k-nearest neighbor algorithm with weighted {RSSI},'' \emph{Procedia Computer
  Science}, vol.~5, pp. 58--65, 2011, publisher: Elsevier.

\bibitem{sahinoglu_ultra-wideband_2008}
Z.~Sahinoglu, S.~Gezici, and I.~Guvenc, ``Ultra-wideband positioning systems,''
  \emph{Cambridge, New York}, 2008.

\bibitem{au_latest_2016}
E.~Au, ``The {Latest} {Progress} on {IEEE} 802.11 mc and {IEEE} 802.11 ai
  [{Standards}],'' \emph{IEEE Vehicular Technology Magazine}, vol.~11, no.~3,
  pp. 19--21, 2016, publisher: IEEE.

\bibitem{hashem_accurate_2021}
O.~Hashem, K.~A. Harras, and M.~Youssef, ``Accurate indoor positioning using
  {IEEE} 802.11 mc round trip time,'' \emph{Pervasive and Mobile Computing}, p.
  101416, 2021, publisher: Elsevier.

\bibitem{gentner_wifi-rtt_2020}
C.~Gentner, M.~Ulmschneider, I.~Kuehner, and A.~Dammann, ``{WiFi}-{RTT} indoor
  positioning,'' in \emph{2020 {IEEE}/{ION} position, location and navigation
  symposium ({PLANS})}, 2020, pp. 1029--1035.

\bibitem{dammann_prospects_2015}
A.~Dammann, R.~Raulefs, and S.~Zhang, ``On prospects of positioning in {5G},''
  in \emph{2015 {IEEE} international conference on communication workshop
  ({ICCW})}, 2015, pp. 1207--1213.

\bibitem{buehrer_collaborative_2018}
R.~M. Buehrer, H.~Wymeersch, and R.~M. Vaghefi, ``Collaborative sensor network
  localization: {Algorithms} and practical issues,'' \emph{Proceedings of the
  IEEE}, vol. 106, no.~6, pp. 1089--1114, 2018, publisher: IEEE.

\bibitem{suryavanshi_direction_2019}
N.~Suryavanshi, K.~V. Reddy, and V.~Chandrika, ``Direction finding capability
  in bluetooth 5.1 standard,'' in \emph{Proceedings {Ubiquitous}
  {Communications} and {Network} {Computing}}, Bangalore, India, Feb. 2019,
  tex.proceedings\_a: UBICNET tex.proceedings: Ubiquitous Communications and
  Network Computing. Second EAI International Conference, Bangalore, India,
  February 8–10, 2019, Proceedings.

\bibitem{noauthor_wi-fi_nodate}
\BIBentryALTinterwordspacing
(2021) \BIBforeignlanguage{en}{Wi-{Fi} {RTT} ({IEEE} 802.11mc)}. [Online].
  Available: \url{https://source.android.com/devices/tech/connect/wifi-rtt}
\BIBentrySTDinterwordspacing

\bibitem{bullmann2020comparison}
M.~Bullmann, T.~Fetzer, F.~Ebner, M.~Ebner, F.~Deinzer, and M.~Grzegorzek,
  ``Comparison of 2.4 {GHz} {WiFi} {FTM}-and {RSSI}-{Based} {Indoor}
  {Positioning} {Methods} in {Realistic} {Scenarios},'' \emph{Sensors},
  vol.~20, no.~16, p. 4515, 2020, publisher: Multidisciplinary Digital
  Publishing Institute.

\bibitem{horn2020doubling}
B.~K. Horn, ``Doubling the {Accuracy} of {Indoor} {Positioning}: {Frequency}
  {Diversity},'' \emph{Sensors}, vol.~20, no.~5, p. 1489, 2020, publisher:
  Multidisciplinary Digital Publishing Institute.

\bibitem{ibrahim2018verification}
\BIBentryALTinterwordspacing
M.~Ibrahim, H.~Liu, M.~Jawahar, V.~Nguyen, M.~Gruteser, R.~Howard, B.~Yu, and
  F.~Bai, ``Verification: {Accuracy} evaluation of {WiFi} fine time
  measurements on an open platform,'' in \emph{Proceedings of the 24th annual
  international conference on mobile computing and networking}, ser. {MobiCom}
  '18.\hskip 1em plus 0.5em minus 0.4em\relax New York, NY, USA: Association
  for Computing Machinery, 2018, pp. 417--427, number of pages: 11 Place: New
  Delhi, India. [Online]. Available:
  \url{https://doi.org/10.1145/3241539.3241555}
\BIBentrySTDinterwordspacing

\bibitem{88W8987DB1x1}
\BIBentryALTinterwordspacing
``{88W8987} {DB} 1x1 802.11ac, {Bluetooth} 5.2.'' [Online]. Available:
  \url{https://www.nxp.com/products/wireless/wi-fi-plus-bluetooth/2-4-5-ghz-dual-band-1x1-wi-fi-5-802-11ac-plus-bluetooth-5-2-solution:88W8987}
\BIBentrySTDinterwordspacing

\bibitem{88MW32XWiFiMicrocontroller}
\BIBentryALTinterwordspacing
``{88MW32X} {Wi}-{Fi}{Microcontroller} {SoC}.'' [Online]. Available:
  \url{https://www.nxp.com/products/wireless/wi-fi-plus-bluetooth/88mw32x-802-11n-wi-fi-microcontroller-soc:88MW32X}
\BIBentrySTDinterwordspacing

\bibitem{WirelessMCUs}
\BIBentryALTinterwordspacing
``Wireless {MCUs}.'' [Online]. Available:
  \url{https://www.cypress.com/products/wireless-mcus}
\BIBentrySTDinterwordspacing

\bibitem{ReleaseESPIDFPrerelease}
\BIBentryALTinterwordspacing
``\BIBforeignlanguage{en}{Release {ESP}-{IDF} {Pre}-release v4.3-beta1 ·
  espressif/esp-idf}.'' [Online]. Available:
  \url{https://github.com/espressif/esp-idf/releases/tag/v4.3-beta1}
\BIBentrySTDinterwordspacing

\bibitem{noauthor_global_nodate}
\BIBentryALTinterwordspacing
(2021) Global {SOC} as a {Service} {Market} {Size}, and {Forecast} to 2028.
  [Online]. Available:
  \url{https://www.quincemarketinsights.com/industry-analysis/soc-as-a-service-market}
\BIBentrySTDinterwordspacing

\bibitem{noauthor_esphome_nodate}
\BIBentryALTinterwordspacing
(2021) \BIBforeignlanguage{en}{{ESPHome}}. [Online]. Available:
  \url{https://esphome.io/index.html}
\BIBentrySTDinterwordspacing

\bibitem{noauthor_tasmota_nodate}
\BIBentryALTinterwordspacing
(2021) Tasmota. [Online]. Available:
  \url{https://tasmota.github.io/docs/About/}
\BIBentrySTDinterwordspacing

\bibitem{noauthor_esp_nodate}
\BIBentryALTinterwordspacing
(2021) {ESP} {Easy}. [Online]. Available:
  \url{https://espeasy.readthedocs.io/en/latest/ESPEasy/AboutUs.html}
\BIBentrySTDinterwordspacing

\bibitem{2pv8-ze59-21}
\BIBentryALTinterwordspacing
V.~Barral, O.~Campos, T.~Domínguez-Bolaño, C.~J. Escudero, and J.~A.
  García-Naya, ``{ESP32S2} {FTM} {Measurements},'' 2021, tex.referencetype:
  data. [Online]. Available: \url{https://dx.doi.org/10.21227/2pv8-ze59}
\BIBentrySTDinterwordspacing

\bibitem{esp32ftmtag}
\BIBentryALTinterwordspacing
V.~Barral. (2021) {ESP32} {FTM} tag source code. [Online]. Available:
  \url{https://github.com/valentinbarral/esp32s2-ftm-tag}
\BIBentrySTDinterwordspacing

\bibitem{esp32ftmanchor}
\BIBentryALTinterwordspacing
------. (2021) {ESP32} {FTM} anchor source code. [Online]. Available:
  \url{https://github.com/valentinbarral/esp32s2-ftm-anchor}
\BIBentrySTDinterwordspacing

\bibitem{rosftmnode}
\BIBentryALTinterwordspacing
------. (2021) {ROS} {FTM} reader source code. [Online]. Available:
  \url{https://github.com/valentinbarral/rosftm}
\BIBentrySTDinterwordspacing

\bibitem{rosftmmsgs}
\BIBentryALTinterwordspacing
------. (2021) {ROS} messages source code. [Online]. Available:
  \url{https://github.com/valentinbarral/rosmsgs}
\BIBentrySTDinterwordspacing

\bibitem{dvorecki2019machine}
N.~Dvorecki, O.~Bar-Shalom, L.~Banin, and Y.~Amizur, ``A machine learning
  approach for wi-fi {RTT} ranging,'' in \emph{Proceedings of the 2019
  {International} {Technical} {Meeting} of {The} {Institute} of {Navigation}},
  Hyatt Regency Reston Reston, Virginia, Jan. 2019, pp. 435 -- 444.

\bibitem{choi2019unsupervised}
J.~Choi, Y.-S. Choi, and S.~Talwar, ``Unsupervised {Learning} {Techniques} for
  {Trilateration}: {From} {Theory} to {Android} {APP} {Implementation},''
  \emph{IEEE Access}, vol.~7, pp. 134\,525--134\,538, 2019.

\bibitem{espresssif_systems_esp32-s2_2021}
\BIBentryALTinterwordspacing
{Espresssif Systems}. (2021) {ESP32}-{S2}. [Online]. Available:
  \url{https://www.espressif.com/en/products/socs/esp32-s2}
\BIBentrySTDinterwordspacing

\bibitem{rosweb}
\BIBentryALTinterwordspacing
{ROS}. (2021) {ROS} {Main} {Site}. [Online]. Available:
  \url{http://www.ros.org}
\BIBentrySTDinterwordspacing

\bibitem{wymeersch2012machine}
H.~Wymeersch, S.~Marano, W.~M. Gifford, and M.~Z. Win, ``A {Machine} {Learning}
  {Approach} to {Ranging} {Error} {Mitigation} for {UWB} {Localization},''
  \emph{IEEE Transactions on Communications}, vol.~60, no.~6, pp. 1719--1728,
  2012.

\bibitem{barral_environmental_2019}
V.~Barral, C.~J. Escudero, J.~A. García-Naya, and P.~Suárez-Casal,
  ``Environmental {Cross}-{Validation} of {NLOS} {Machine} {Learning}
  {Classification}/{Mitigation} with {Low}-{Cost} {UWB} {Positioning}
  {Systems},'' \emph{Sensors}, vol.~19, no.~24, p. 5438, 2019, publisher:
  Multidisciplinary Digital Publishing Institute.

\bibitem{pelikan1999boa}
M.~Pelikan and D.~E. Goldberg, ``{BOA}: {The} {Bayesian} optimization
  algorithm,'' in \emph{in {Proc}. {Genetic} and}, 1999, pp. 525--532.

\bibitem{moisen2008classification}
\BIBentryALTinterwordspacing
G.~G. Moisen, ``\BIBforeignlanguage{en}{Classification and regression trees},''
  \emph{\BIBforeignlanguage{en}{In: Jørgensen, Sven Erik; Fath, Brian D.
  (Editor-in-Chief). Encyclopedia of Ecology, volume 1. Oxford, U.K.: Elsevier.
  p. 582-588.}}, pp. 582--588, 2008. [Online]. Available:
  \url{https://www.fs.usda.gov/treesearch/pubs/30645}
\BIBentrySTDinterwordspacing

\bibitem{vapnik1995nature}
\BIBentryALTinterwordspacing
V.~Vapnik, \emph{\BIBforeignlanguage{en}{The {Nature} of {Statistical}
  {Learning} {Theory}}}, 2nd~ed., ser. Information {Science} and
  {Statistics}.\hskip 1em plus 0.5em minus 0.4em\relax New York:
  Springer-Verlag, 2000. [Online]. Available:
  \url{https://www.springer.com/gp/book/9780387987804}
\BIBentrySTDinterwordspacing

\bibitem{drucker1997support}
H.~Drucker, C.~Burges, L.~Kaufman, A.~Smola, and V.~Vapnik, ``Support vector
  regression machines,'' \emph{Adv Neural Inform Process Syst}, vol.~28, pp.
  779--784, Jan. 1997.

\bibitem{rasmussen2003gaussian}
\BIBentryALTinterwordspacing
C.~E. Rasmussen, ``Gaussian {Processes} in {Machine} {Learning},'' in
  \emph{Advanced {Lectures} on {Machine} {Learning}: {ML} {Summer} {Schools}
  2003, {Canberra}, {Australia}, {February} 2 - 14, 2003, {Tübingen},
  {Germany}, {August} 4 - 16, 2003, {Revised} {Lectures}}, O.~Bousquet, U.~von
  Luxburg, and G.~Rätsch, Eds.\hskip 1em plus 0.5em minus 0.4em\relax Berlin,
  Heidelberg: Springer Berlin Heidelberg, 2004, pp. 63--71. [Online].
  Available: \url{https://doi.org/10.1007/978-3-540-28650-9_4}
\BIBentrySTDinterwordspacing

\bibitem{schmidhuber2015deep}
\BIBentryALTinterwordspacing
J.~Schmidhuber, ``Deep {Learning} in {Neural} {Networks}: {An} {Overview},''
  \emph{Neural Networks}, vol.~61, pp. 85--117, Jan. 2015, arXiv: 1404.7828.
  [Online]. Available: \url{http://arxiv.org/abs/1404.7828}
\BIBentrySTDinterwordspacing

\bibitem{CP2102NGQFN28USBXpressUSB2021}
\BIBentryALTinterwordspacing
(2021) {CP2102N}-{GQFN28} {USBXpress} {USB} {Bridge} - {Silicon} {Labs}.
  [Online]. Available:
  \url{https://www.silabs.com/interface/usb-bridges/usbxpress/device.cp2102n-gqfn28}
\BIBentrySTDinterwordspacing

\bibitem{pincheiraCosteffectiveIoTDevices2021}
M.~Pincheira, M.~Vecchio, R.~Giaffreda, and S.~S. Kanhere, ``Cost-effective
  {{IoT}} devices as trustworthy data sources for a blockchain-based water
  management system in precision agriculture,'' \emph{Computers and Electronics
  in Agriculture}, vol. 180, p. 105889, Jan. 2021.

\bibitem{bougueraEnergyConsumptionModel2018}
T.~Bouguera, J.-F. Diouris, J.-J. Chaillout, R.~Jaouadi, and G.~Andrieux,
  ``Energy {{Consumption Model}} for {{Sensor Nodes Based}} on {{LoRa}} and
  {{LoRaWAN}},'' \emph{Sensors}, vol.~18, no.~7, p. 2104, Jul. 2018.

\end{thebibliography}
\bibliographystyle{IEEEtran}

\vskip -2.4\baselineskip plus -1fil
\begin{IEEEbiography}
	[{\includegraphics[width=1in,height=1.25in,clip,keepaspectratio]{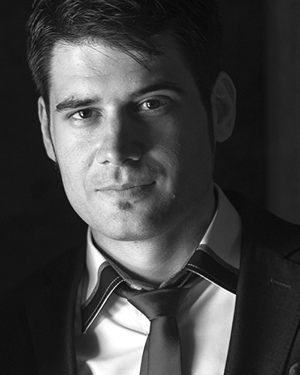}}
	]{Valentín Barral Vales} works as an associated researcher at University of A Coruña, where he took his degree as a computer engineer and his Ph.D. with a thesis titled: \textit{Ultra Wideband location in scenarios without clear line of sight}. His research field is indoor localization using radio technologies, a field in which he has published several works since 2012. He has also participated in many national and European projects, always related to indoor localization in complex environments.
\end{IEEEbiography}
\vskip -2.4\baselineskip plus -1fil
\begin{IEEEbiography}[{\includegraphics[width=1in,height=1.25in,clip,keepaspectratio]{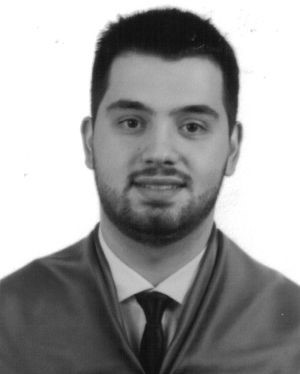}}
	]{Omar Campos Fernández} received the B.S degree in Electronics Engineering from the University of A Coruña (UDC), A Coruña, Spain in 2019 and the M.Sc in Electronics Systems Engineering from the University Carlos III of Madrid (UC3M), Madrid, Spain in 2020. His research interests include indoor localization systems and wireless sensor networks.
\end{IEEEbiography}
\vskip -2.4\baselineskip plus -1fil
\begin{IEEEbiography}[{\includegraphics[width=1in,height=1.25in,clip,keepaspectratio]{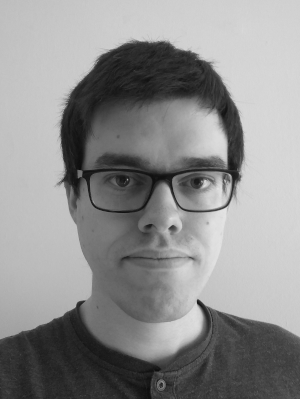}}
	]{Tomás Domínguez-Bolaño} received the B.S degree in Computer Engineering and the Ph.D. in Computer Engineering (with the distinction ``Doctor with European Mention'') from the University of A Coruña, A Coruña, Spain, in 2014 and 2018, respectively. Since 2014 he has been with the Group of Electronics Technology and Communications. In 2018 he was a Visiting Scholar with Tongji University, Shanghai, China. He is an author of more than 15 papers in peer-reviewed international journals and conferences. He was awarded with a predoctoral grant and two research-stay grants. His research interests include channel measurements, parameter estimation and modeling and experimental evaluation of wireless communication systems.
\end{IEEEbiography}
\vskip -2.4\baselineskip plus -1fil
\begin{IEEEbiography}
	[{\includegraphics[width=1in,height=1.25in,clip,keepaspectratio]{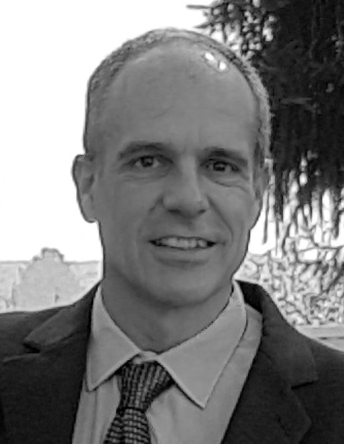}}
	]{Carlos J. Escudero} received the Ph.D. degree in Computer Science from University of A Coruña (UDC), Spain in 1998. He is a Full Professor at the University of A Coruña (UDC), Spain, in the area of Signal Theory and Communications. Since 1992, he has been a Faculty Member with the Department of Computer Engineering (UDC), where he became Associate Professor in 1998. He was Secretary of the Department, Vice-Dean of the Faculty of Informatics, Head of the Department of Electronics and Systems  and, since January 2016 to present, Deputy Rector of the UDC for Information and Communication Technologies (University CIO). His research focuses on applications of signal processing in communications, wireless sensor networks, and indoor location systems. Proof of this are some of the merits alleged in the curriculum: publications in the most prestigious journals and books, many national and international collaborations with renowned researchers, dozens of papers / publications in national and international conferences, highlighting the most significant in the curriculum, participation in more than 25 competitive research projects in which he has been the principal investigator on 7, responsible for 20 contracts with companies, author of two patents, and founding member of two spin-off companies.
\end{IEEEbiography}
\vskip -2.4\baselineskip plus -1fil
\begin{IEEEbiography}[
	{\includegraphics[width=1in,height=1.25in,clip,keepaspectratio]{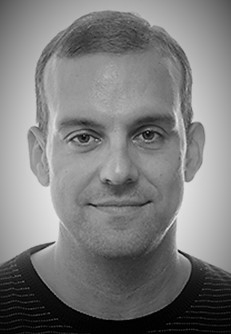}}
	]{José A. García-Naya} (Senior Member, IEEE) received the M.Sc. and Ph.D. degrees in computer engineering from the University of A Coruña (UDC), Spain, in 2005 and 2011, respectively. He is with the CITIC Research Center and with the Group of Electronics Technology and Communications, both at UDC, where he is Associate Professor. He is the coauthor of more than 120 peer-reviewed papers in journals and conferences. He was member of the research team in more than 40 research projects funded by public organizations and private companies, being principal investigator in two of them. His research interests focus on wireless engineering, with special emphasis on experimental evaluation, including wireless channel characterization, high-mobility vehicular transportation, time-modulated arrays; location systems, and IoT.
\end{IEEEbiography}

\end{document}